\documentclass{aa}
\usepackage{graphics} 
\usepackage{epsfig}

\def\ergsec{\hbox{erg s$^{-1}$ }}
\def\ergcm{\hbox{erg cm$^{-2}$ s$^{-1}$ }}
\def\HI{\hbox{H\,{\sc i}}}
\def\HII{\hbox{H\,{\sc ii}}}

\begin{document}
   \thesaurus{3;(11.09.1 NGC 253; 11.19.2; 11.19.3; 13.25.2)}
   \title{
          X-ray observations of the starburst galaxy NGC~253: \\
          I. Point sources in the bulge, disk and halo 
             \thanks{
             based partially on observations performed at the European
             Southern Observatory, La Silla, Chile}
          }
  
   \author{A.~Vogler\inst{1,2} \and W.~Pietsch\inst{1}}

   \offprints{A.~Vogler (ajv@mpe.mpg.de)}
 
   \institute{Max-Planck-Institut f\"ur extraterrestrische Physik,
              Gie\ss enbachstra\ss e, D--85740 Garching,\\
              Federal Republic of Germany
   \and       CEA/Saclay, DAPNIA, Service d'Astrophysique,
              L'Ormes des Merisiers, B\^at. 709,\\
              F--91191 Gif-sur-Yvette, France}

   \date{Received date; accepted date}
   \titlerunning{X-ray point sources in the bulge, disk and halo of NGC~253}

   \maketitle

%   \markboth{A. Vogler \& W. Pietsch: X-ray point sources in the bulge, disk 
%                                      and halo of NGC~253}{}

   \begin{abstract}
We report the results of a deep spatial, spectral, and timing analysis of 
ROSAT HRI and PSPC observations of the edge-on starburst galaxy NGC~253. 
In this first paper, point-like X-ray sources detected within the galaxy 
and in the field are discussed. The sources are characterized by their 
X-ray properties (including comparisons with results from the {\it Einstein} 
and ASCA satellites), by correlations
with other wavelength and some optical spectroscopic follow up observations.

In total, 73 X-ray sources have been collected in the NGC~253 field, 
32 of which are 
associated with the disk of the galaxy. Though 27 of these disk sources are
detected with the HRI (some being resolvable with the 
PSPC), the remaining 5 PSPC-only detected sources are likely not to be
real point sources, being instead due to fluctuations within the X-ray 
structure of the disk.
The source close to the center of the galaxy is extended 
($L_{\rm x}\sim 1\times 10^{39}$~erg~s$^{-1}$
in the ROSAT 0.1--2.4~keV band), and is most likely associated with the nuclear 
starburst activity. The remaining sources have
luminosities ranging from $7\times 10^{36}$~erg~s$^{-1}$ 
to $3.0\times 10^{38}$~erg~s$^{-1}$, yielding an integrated point source
luminosity of $1\times 10^{39}$~erg~s$^{-1}$. 
The brightest point-like source is located $\sim 20''$ south of the 
nucleus, at the border of a plume of diffuse X-ray emission. Its high
X-ray luminosity, time variability and hard spectrum make it a good 
candidate for a black hole X-ray binary. 

Including four {\it Einstein} detections of X-ray transients
the number of point-like X-ray sources in NGC~253 increases to 30 sources, 
13 of which are time variable. 
These time variable sources are all brighter than
$5\times 10^{37}$~erg~s$^{-1}$ and most likely represent X-ray binaries 
radiating close to or at the Eddington limit. Besides the nuclear source
there is only one source above this luminosity that shows no time
variability and therefore may represent a young supernova or extremely
bright supernova remnant, or an unresolved cluster of several X-ray sources.
The point source population of NGC~253 is compared to that of other galaxies, 
and it is
shown that the luminosity distribution matches ROSAT results
obtained for M~31 and M~33. 

The halo of NGC~253 is filled with diffuse, filamentary X-ray emission. 
Seven sources 
are located (or projected) in this diffuse emission region. Time variability
arguments, together with optical identifications, are put forward to
explain 4 sources as background objects, the other 3 sources 
likely being spurious detections caused by
local enhancements in the diffuse emission of the halo of NGC~253. 
The diffuse X-ray emission components of NGC~253 will be discussed in a 
separate paper. 

The sources detected in the field outside the disk of NGC~253 cover a flux range
from $(9 - 300)\times 10^{-15}$\ergcm in the 0.1--2.4 keV band. 
None of the sources in the field correlate with published lists of 
globular cluster candidates. Optical
counterparts are proposed for 27 of them, and a few also correlate with radio
sources. While two sources are identified as foreground stars, the
remaining ones are mostly background active galactic nuclei.

   \end{abstract}
\section{Introduction}

NGC~253 is a nearby, almost edge-on spiral galaxy. Some physical 
parameters of the  
galaxy, such as distance, inclination or $D_{25}$ diameter, are summarized in 
Table~\ref{parameters}. The galaxy is classified as a starburst galaxy with
a very high level of nuclear starburst activity (cf. Carral et al. 1994 
for an overview). 
Detailed X-ray observations of the galaxy have been carried out in the past
with the {\it Einstein} satellite 
(Fabbiano \& Trinchieri 1984, Fabbiano 1988). More recent X-ray observations
obtained with the ASCA satellite are reported by Ptak et al. (1997). 

   \begin{table}
      \caption{Parameters of NGC~253.}
         \label{parameters}
         \begin{flushleft}
         \begin{tabular}{lrr}
            \hline
            \noalign{\smallskip}
 & &Ref.  \\
            \noalign{\smallskip}
            \hline
            \noalign{\smallskip}
%Type & SAB(s)c & $^\ast$ \\
Type &  Sc & $^\ast$ \\
            \noalign{\smallskip}
Assumed distance & 2.58 Mpc  & $^{\sharp}$ \\
& (hence 1$'\cor750$~pc) & \\
            \noalign{\smallskip}
Position of &  $\alpha_{2000}= 0^{\rm h}~47^{\rm m}~33\fs3$& $^\dagger$\\
center (2000.0) & $\delta_{2000}=-25\degr 17\arcmin 18''$ & \\
            \noalign{\smallskip}
$D_{25}$ & 25\farcm 4 & $^\ast$ \\
            \noalign{\smallskip}
Corrected $D_{25}$ & 18\farcm 8 & $^\ast$ \\
            \noalign{\smallskip}
Axial ratio & 0.23 & $^\ast$ \\
            \noalign{\smallskip}
Position angle & 52\degr & $^\ddagger$ \\
            \noalign{\smallskip}
Inclination & 86\degr & $^\ast$ \\
            \noalign{\smallskip}
Galactic foreground $N_{\rm H}$ &1.3$\times10^{20}$~cm$^{-2}$ &$^\clubsuit$ \\
            \noalign{\smallskip}
            \hline
            \noalign{\smallskip}
         \end{tabular}
         \end{flushleft}
{
References: \\
$^\ast$ \ \ Tully (1988)\ \\
%$^\ast$ \ \ de Vaucouleurs et al. (1991)\ \\
$^{\sharp}$ \ \ Puche \& Carignan (1988)\\
$^\dagger$ \ \ Forbes et al. (1991)\\
%$^\heartsuit$  \ Tully (1988)\\ 
$^\ddagger$ \ \ SIMBAD data base, operated at CDS, Strasbourg, France \\
$^\clubsuit$ \  Dickey \& Lockman (1990)\\
}
   \end{table}

The {\it Einstein} HRI and IPC measurements 
established a total X-ray luminosity of 
$L_{\rm x} = 3 \times 10^{39}$~erg~s$^{-1}$ and 
$L_{\rm x} = 4 \times 10^{39}$~erg~s$^{-1}$, respectively, for the disk of
NGC~253 (0.2--4~keV {\it Einstein} 
band, corrected for Galactic foreground absorption). Eight point-like sources 
with luminosities 
$\ga 6 \times 10^{37}$~erg~s$^{-1}$ were detected
in the bulge and disk of the 
galaxy, with an integrated point source luminosity of
$1 \times 10^{39}$~erg~s$^{-1}$. The brightest point source, $\sim 
20''$ south of the center of the galaxy, reached 
$L_{\rm x} \sim 3 \times 10^{38}$~erg~s$^{-1}$, and
X-ray emission in the inner bulge of the galaxy was discernible
with $L_{\rm x} = 1 \times 10^{39}$~erg~s$^{-1}$. A plume-like, diffuse
X-ray emission feature (`jet') 
protrudes along the minor axis from the nucleus towards the south east, with
a luminosity of
$\sim 3 \times 10^{38}$~erg~s$^{-1}$. Fabbiano \& Trinchieri (1984)
attributed this emission to a hot 
($T\sim 10^{(7.5)}$~K), gaseous component of the interstellar medium,
connected to the nuclear starburst activity. Additional diffuse X-ray 
emission from the hot component was detected both, in the disk ($L_{\rm x} = 1 
\times 10^{39}$~erg~s$^{-1}$) and to the north western side of the 
halo ($L_{\rm x} = 1 \times 10^{39}$~erg~s$^{-1}$) (Fabbiano 1988). 

Contrary to the {\it Einstein} data, ASCA observations allow detailed
spectral analysis
of the integral NGC~253 emission ($6\times 10^{39}$~erg~s$^{-1}$
in the 0.5--10~keV ASCA band). However, 
it is difficult to disentangle the different point sources 
due to the large ASCA point spread function (PSF) of
$\sim 3'$ at Full Width Half Maximum (FWHM). %{\bf ref?}

We performed deep X-ray observations of NGC~253 with the ROSAT high resolution
imager (HRI) and position sensitive proportional counter (PSPC). 
A description of the satellite and the detectors on board is given in 
Tr\"umper (1983) and Pfeffermann et al. (1987). 
Whereas the ROSAT HRI allows good spatial 
separation of point sources from diffuse emission, 
due to its narrow PSF ($\sim 5''$
FWHM), with the PSPC (high sensitivity in the 0.1 -- 2.4 keV band, 
$\sim 25''$ FWHM for the 0.5--2.0~keV band),
one can perform spectral analysis 
for bright point-like sources and diffuse emission components, and establish 
X-ray colors (hardness ratios).

\begin{figure*}
\includegraphics[bb=54 254 508 671,width=13cm,clip=true]{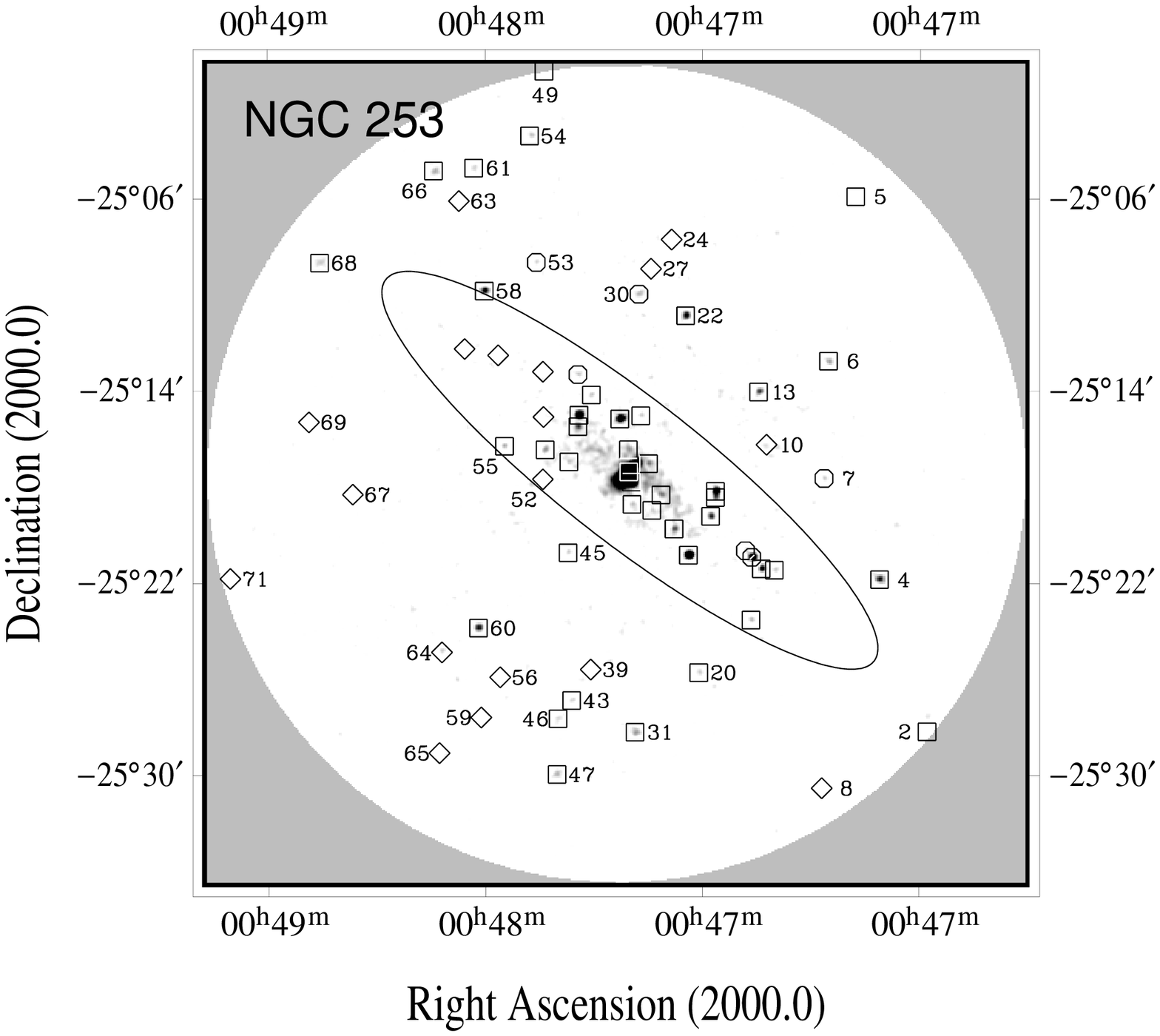}
\caption{
ROSAT HRI image of the NGC~253 pointing for the full HRI field of 
view. The image has been formed
with a binsize of 5$''$ and smoothed with a Gaussian filter of $12''$ FWHM. 
Detected X-ray sources (cf. Sect.~\ref{howhard} and Table~3)
are indicated by squares (sources
detected with the HRI and PSPC), hexagons (sources only detected with the HRI) 
or diamonds (sources only detected with the PSPC). The $D_{25}$ ellipse of 
NGC~253 has been sketched, the center of the galaxy coincides with the 
position of the central source. 
Sources outside the area covered by the disk of 
NGC~253 have been enumerated in this figure, the others are enumerated in 
Fig.~2. The right ascension and declination are given for J2000
        }
\end{figure*}

\begin{figure*}
\includegraphics[bb=54 254 508 671,width=13cm,clip=true]{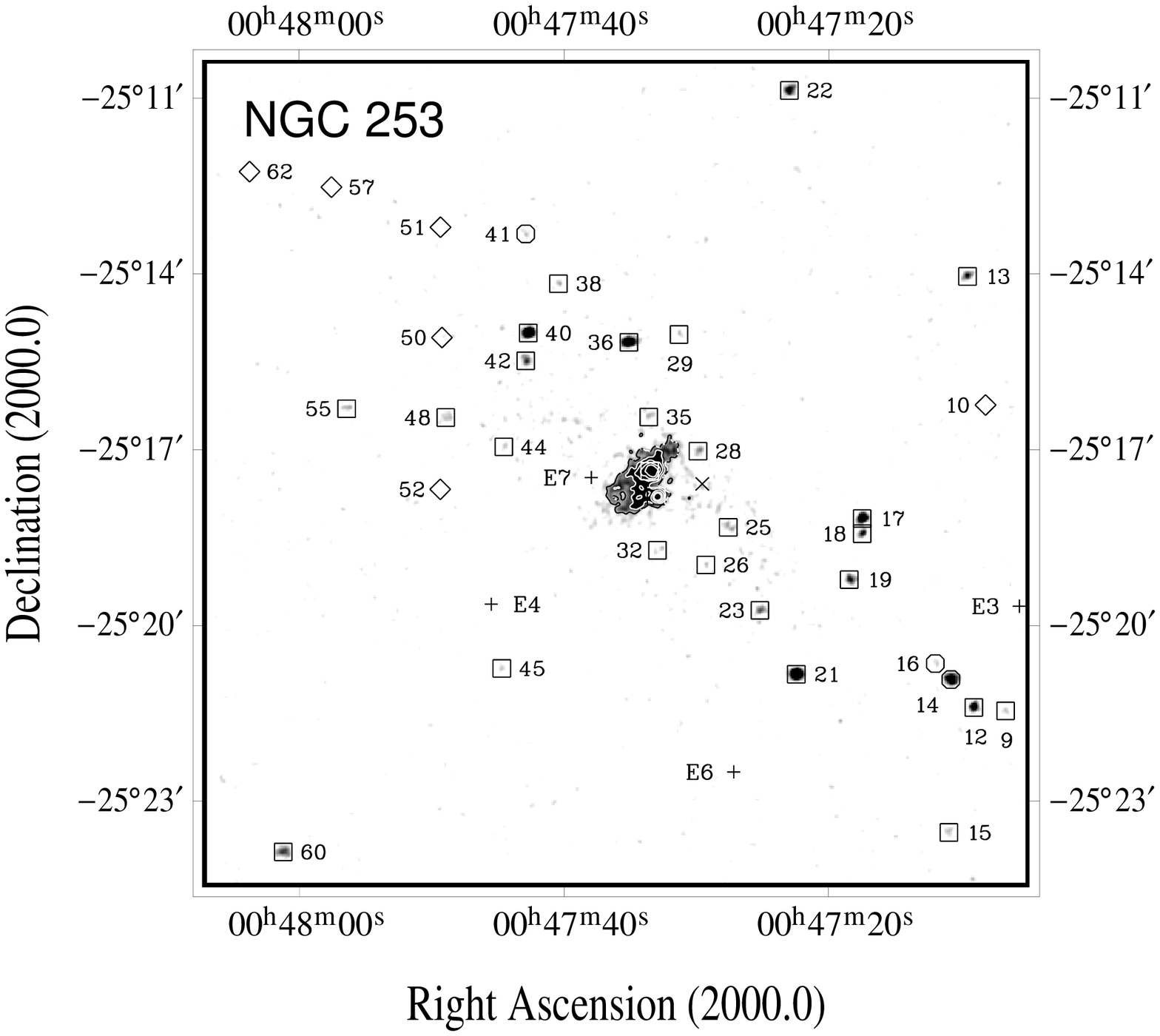}
\caption{
ROSAT HRI image for the inner 14$'$ field of the pointing. The image
has been formed with a binsize of 1$''$ and smoothed with a Gaussian filter
of $4\farcs 7$. Sources have been 
marked according to Fig.~1. All sources but the central one (X34) and a bright
source 20$''$ to the south of X34
(X33) have been enumerated according to Table~3. The cross ($\times$) between
the sources X25 and X28 marks the position of the SN 1940 E.
The positions of the ROSAT sources X17, X21, X33 and X36 coincide with the
positions of the {\it Einstein} sources E2, E5, E8 and E1, respectively.
The {\it Einstein} sources E3, E4, E6 and E7 remained undetected during the
ROSAT observations, and the positions of these sources have been marked with
crosses ($+$). 
To make the structure of the over-exposed central region of the galaxy
visible, contours (4, 8, 15, 27 and
$40\times 4.4 \cdot 10^{-3}$~cts~s$^{-1}$~arcmin$^{-2}$)
have been over-plotted in the over-exposed area. 
Fig.~3 shows a close up view of the over-exposed region
        }
\end{figure*}

In this paper, we make use of the high collecting power of the ROSAT 
telescope/detector
systems to derive a detailed point source catalog for the NGC~253 field 
with special emphasis on the sources in the bulge, 
disk and halo of NGC~253 (Sects. 2 and 3). In Sect. 4, the nature of the
detected point-like sources is discussed and the source catalog is
compared to previous results regarding X-ray point sources in NGC~253 and 
other galaxies and to results from other wavelengths. Sect. 5 gives a 
summary of the results. X-ray sources in the
field of NGC~253, together with identifications proposed from cross-correlations
with optical and radio catalogs and optical follow-up spectroscopy are 
discussed in the appendix.

In a complementary paper (Pietsch et al. 1998a, hereafter PEA), we investigate 
the diffuse X-ray emission features in the bulge, disk and halo of NGC~253,
after subtracting the point sources.

\section{Observations and data analysis}
\label{observations}
NGC~253 was observed with the ROSAT HRI and PSPC for 57.7~ks 
and 22.8~ks, respectively.
Both the HRI and PSPC observations are each spread over approximately 3.5 years 
and consist of 6 HRI and 2 PSPC observation blocks, each subdivided
into several observation intervals (OBIs). The
date and integration times for the different  
observation blocks are listed in Table~\ref{group}.

   \begin{table*}
      \caption{ROSAT observations of NGC~253.}
         \label{group}
         \begin{flushleft}
         \begin{tabular}{lcccr}
            \hline
            \noalign{\smallskip}
   & Det. & Date   &rel. time in days        & $t_{\rm int}$\\
            \noalign{\smallskip}
            \hline
            \noalign{\smallskip}
1           
 & HRI  &  08 NOV -- 10 DEC 1991  & $\phantom{.0}$0 -- 0.9 & 3.1~ks\\
\noalign{\smallskip}
2           
 & PSPC &  25 DEC -- 31 DEC 1991  & 11.9 -- 15.9  & 11.7~ks\\
\noalign{\smallskip}
3 
 & PSPC &  03 JUN -- 05 JUN 1992  & 125.7 -- 126.6 & 11.2~ks\\
\noalign{\smallskip}
4
 & HRI  & 05 JUN -- 06 JUN 1992   & 127.1 -- 127.5 & 15.1~ks\\
\noalign{\smallskip}
5
 & HRI     & 06 JUN -- 07 JUN 1992  & 127.7 -- 127.9 & 10.1~ks\\
\noalign{\smallskip}
6
 & HRI     & 03 JAN -- 07 JAN 1995  & 791.6 -- 794.5 & 9.5~ks\\
\noalign{\smallskip}
7
 & HRI     &  13 JUN -- 15 JUN 1995  & 905.5 -- 906.9 & 13.9~ks\\
\noalign{\smallskip}
8
 & HRI     &  05 JUL  -- 07 JUL 1995  & 921.2 -- 922.7 & 5.9~ks\\
\noalign{\smallskip}
\hline
         \end{tabular}
         \end{flushleft}
   \end{table*}

The data reduction was performed with the ESO-MIDAS/EXSAS 
(ESO-MIDAS 1997, Zimmermann et al. 1997) software package.

\subsection{Attitude corrections} 
Attitude solutions of ROSAT pointings used by the Standard Analysis 
Software System (SASS, Voges 1992) to produce event files are known to
produce residual integral errors of the order of 6$''$ (boresight error)
for an observation, and -- due to short term fluctuations -- systematically 
broaden the PSF. To improve on the solution we adopted two (subsequent)
techniques.

Firstly, for the PSPC observations the positions
of point sources during the first and second PSPC observation block
were compared, and the two blocks -- using different guide star patterns 
for the attitude solution -- were aligned. For the HRI, to improve the 
intrinsically narrower PSF, all observation intervals were aligned with 
respect to the first one by computing the centroid for 8 bright point-like
X-ray sources visible in each OBI. No offset was seen to exceed 4$''$.
 
In a second step, the positions of 13 possible optical counterparts 
(derived from the ROE finding charts, Irwin et al. 1994), coinciding to 
within $6''$ with point source HRI error 
circles, were used to determine the systematical 
offset of the preliminary attitude solution. 
The HRI observations were corrected for this
offset (translation of 1\farcs9 and 1\farcs1 to the E and N, respectively). The
PSPC observations were co-aligned
(3\farcs0 and 3\farcs6 to the E and N, respectively, 
counterclockwise rotation of 0.27$^\circ$) with the HRI solution.
All X-ray source positions given in this paper have already been 
transformed into the sky coordinate system. The systematical error of the 
final source positions is determined as the residual error of the
transformations (2\farcs5).

\subsection{Image generation}
\label{images}

\subsubsection{Images of the HRI observations}

To reduce the background due to UV emission and cosmic rays, 
HRI images were integrated using raw channels 2--8.
An 35$'$ image (0.1--2.4~keV band) was constructed 
with a bin size of
2\farcs5 and smoothed with a Gaussian filter of 12$''$ FWHM (Fig.~1).
The optical extent of the galaxy is indicated by  
the $D_{25}$ ellipse. A close-up view of the inner
14$'$ field of the HRI (Fig.~2) has been formed 
with a bin size of 1$''$ and smoothed with a filter of 5$''$ FWHM,
corresponding to the FWHM of the on-axis HRI PSF.
Within a radius of $7'$ from the center of the HRI, the PSF does not
deteriorate significantly. 
To investigate the detailed structure of the central 2$'$ region of the 
galaxy (over-exposed in Figs.~1 and 2), we overlaid X-ray contours
over a greyscale plot of the central area of the image used for Fig. 2 
(Fig.~3).

\subsubsection{Image of the PSPC observations} \label{smooth}

A 0.1--2.4~keV PSPC image (Fig.~4) was
constructed by the superposition of sub-images in 8 standard bands
(R1 to R8, cf. Snowden et al. 1994).
Each sub-image has been corrected for exposure, deadtime, and vignetting, and 
the sub-images have been smoothed with a Gaussian filter
corresponding to the on-axis PSF of the energy band (FWHM
ranging from 52$''$ to 24$''$ for the lowest to the highest energy band).
To make the full dynamic
range of the PSPC data visible in one image, a greyscale 
representation is chosen running several times from bright to dark. 

\subsection{Constructing a combined HRI/PSPC point source catalog}
\subsubsection{HRI point source detection}
\label{searchhri}
Point sources were searched for in the full HRI field of view
with the EXSAS local detect, map detect and
maximum likelihood algorithms (Zimmermann et al. 1997),
using images
of pixel size 5$''$. To reduce the background due to
UV emission and cosmic rays only those events
detected in HRI raw channels 2--8
were used. Sources with a detection likelihood 
$\gid 8$ were accepted. Maximum likelihood values ($L$)
can be converted into probabilities ($P$) through $P=1-e^{-L}$, thus
$L=8$ corresponds to 3.6 Gaussian sigma significance
(cf., e.g. Cruddace et. al 1988). 

Within a field centered on the nucleus of NGC~253 and extending
$\sim 6'$ along the major and $\sim 2\farcm5$ along the minor axis, 
point sources are seen embedded in extended emission structures, 
not resolved by the HRI. Since
the background map in this region does not follow the filamentary 
diffuse structures, a maximum likelihood algorithm,
comparing the local X-ray brightness with the local value of the background 
map, might
pick up extended emission regions as point sources. To avoid this, we
only accepted sources detected by the local detect algorithm in regions
of enhanced diffuse emission. 
The local detect algorithm looks for gradients in the image 
with help of a sliding box method and adapts better to local changes 
in the background. One exception was within a 
bright part of the southeastern extension of the central 
diffuse X-ray emission. Here, we manually 
excluded a spurious source detected by the local detect algorithm at a 
position $\sim 15''$ east of the bright source close to the nucleus. 
Our final HRI source list comprises a total
of 49 sources in the HRI field of view.

\label{hrima}

\subsubsection{PSPC point source detection}
\label{searchpspc}
Source positions and count rates were
calculated for the inner $42'\times42'$ PSPC field  
in the five standard energy bands
`broad' (0.11--2.40~keV), `soft' (0.11--0.41~keV), `hard' (0.52--2.01~keV),
`hard1' (0.52--0.90~keV), and `hard2' (0.91--2.01~keV). All images used for
the source detection have a bin size of 5$''$. As with the HRI, 
the EXSAS local detect, map detect, and
maximum likelihood algorithms were applied to each energy band. 
Sources with a likelihood $\gid$ 9 ($3.8\sigma$)
were accepted, and the source lists within each of the
different energy bands merged, assuming that 
detected source positions were
identical if their separation is less than 3 times the statistical position 
error. The final source position was taken from the energy band
in which the source was found with the highest likelihood.
One PSPC source was located in between two HRI sources (X43 and X46), 
separated by $\sim 1'$. The fact that this PSPC source was flagged as extended
(extent $\sim 1'$ FWHM) by the detection algorithms strongly suggested
that it represents the combined unresolved emission from the two
HRI sources, and this source therefore, was removed from the PSPC source list.
Later, when merging 
the HRI and PSPC lists, the PSPC count rates for X43 and X46 were 
calculated with fixed positions suggested by the HRI.

\begin{figure}
\includegraphics[width=8.8cm,clip=true]{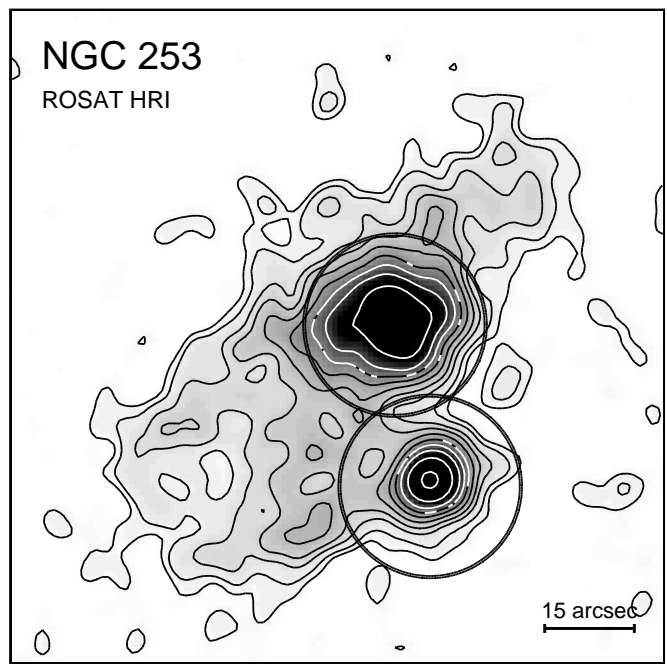}
\caption{
HRI image of the central region of NGC~253. The image gives the inner field 
of Fig.~2. The contours of the same image have been superposed.
Contour levels are at 3, 4, 6, 8, 11, 15, 20, 27, 40, 50 and
$70\times 4.4 \cdot 10^{-3}$~cts~s$^{-1}$~arcmin$^{-2}$. The circles, 
each having a radius of $15''$, correspond to the extraction radius 
for the surface brightness profiles (cf. Sect.~\ref{verybright} and 
Fig.~\ref{radial}), and
mark the positions of X34 (central source) and X33 
        }
\end{figure}

\begin{figure*}
\includegraphics[bb=54 254 508 671,width=13cm,clip=true]{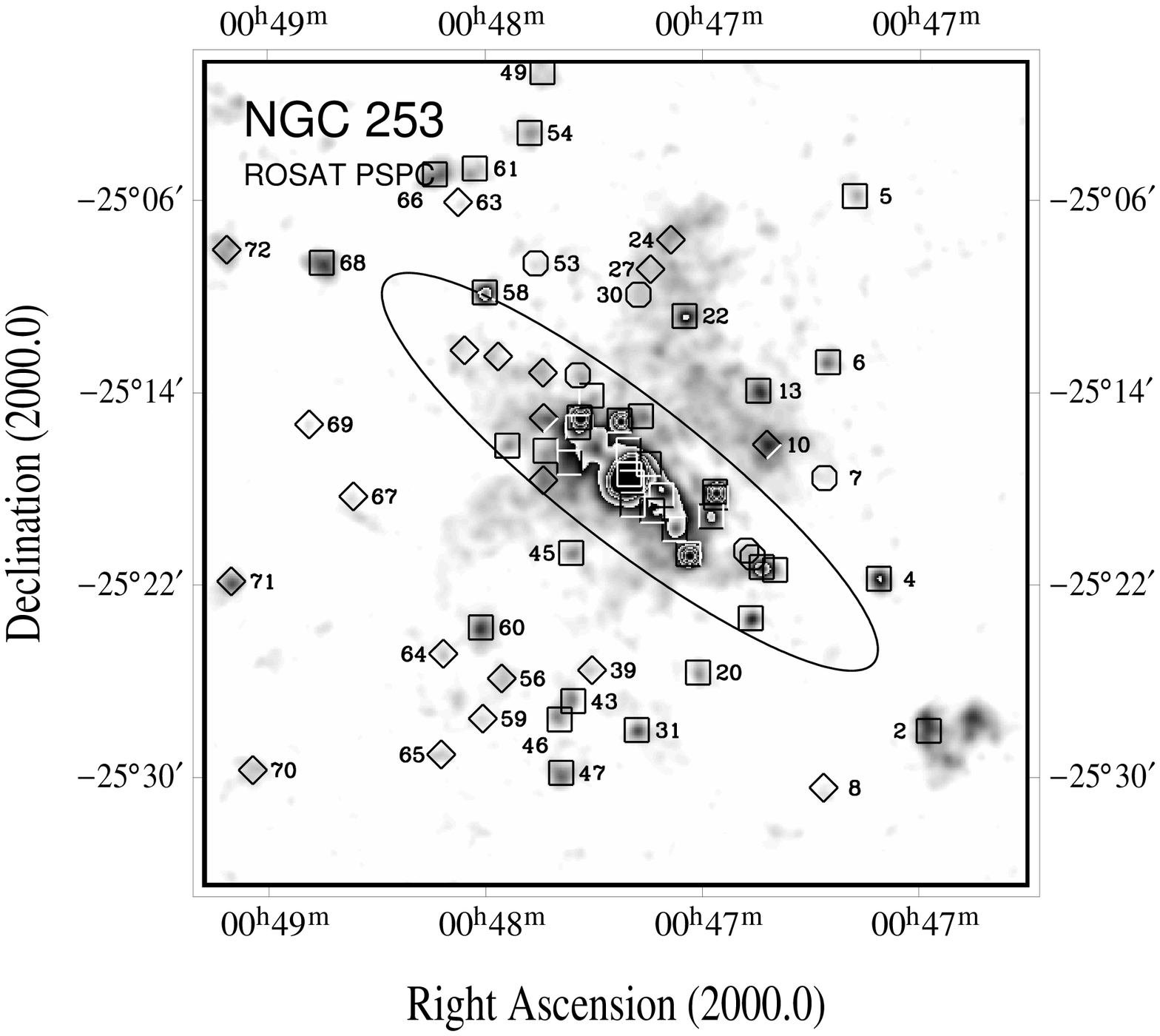}
\caption{
ROSAT PSPC broad band image of the NGC~253 pointing for the same
field as presented in Fig.~1. The image has been formed
with a binsize of $5''$ and smoothed as described in Sect.~\ref{smooth}. The
greyscale representation runs several times from bright to dark to make
the full dynamical range of the data visible. The $D_{25}$ ellipse of NGC~253
has been sketched, sources have been marked according to Fig.~1 and Table~3
        }
\end{figure*}

The energy resolution of the \label{hrrr} 
PSPC detector allows the calculation of
`X-ray colors' for the X-ray sources, the hardness ratios HR1 and HR2. 
The ratios are defined on the basis of the net counts
in the soft, hard, hard1, and hard2 bands. By definition,  
HR1 = (hard--soft)/(hard+soft), and HR2 = (hard2--hard1)/(hard2+hard1)
(soft here means the counts in the soft band etc.). 

\label{pspcma}

\subsubsection{Combined HRI/PSPC point source catalog} \label{howhard}
\label{hriandpspc}

From the HRI and PSPC catalogs, a combined point source 
catalog (Table~3) was constructed. Column 1 gives the
source numbers, that are used to identify sources in Figs.~1 to 4.
The R\,XJ name (following the naming convention for ROSAT sources according
to Zimmermann et al. 1997) is contained in col. 2. Columns 3 and 4 give
the source positions (right ascension and declination, equinox 2000.). 
Sources with spatial separations between the HRI and PSPC positions
smaller than the sum of the 90\% 
error radii were taken as identical
(the `detector flag' in col. 5 has the entry `B' for both). 
If a source existed only in the PSPC or HRI source list, an
entry `P' (PSPC) or `H' (HRI), is given in col. 5. 
Sources have the entry {\it confused} (`C') in col. 6 if the PSPC hard band 
image shows extended emission surrounding the source, or the HRI image
resolves two point sources separated by a distance smaller than the PSPC hard 
band PSF. For confused sources, the position information was always 
taken from the HRI source list. For other 
`B' sources, the position information with the smaller statistical position
error was selected. The
position errors, including a 2\farcs5 systematical error, are given in col. 7.
The likelihood for source detections is displayed in cols. 8 (HRI) and 9
(PSPC). 

\label{ka}
Columns 10 and 11 give the net counts of the X-ray sources,
cols. 12 and 13 the count rates, corrected for exposure, deadtime, and 
vignetting. In the
case of HRI non-detections at PSPC source positions, 
HRI counts were calculated at the position suggested by the PSPC. 
To do so, HRI counts were extracted with a cut radius of 1.5 times
the local FWHM of the HRI PSF at the PSPC source position, and background 
counts subtracted determined at that position in the EXSAS background
maps. For sources with detection likelihoods $<8$, upper limits, at a 2$\sigma$
confidence level, are given. PSPC counts for isolated sources 
were determined using the source detection results.  For confused PSPC
sources with nearby 
point sources (separation below the FWHM of the hard band PSF),
a `multi source fit' technique (cf. Zimmermann et al. 1997) was used with 
source positions fixed according to the HRI detections.
For PSPC confused sources without nearby point sources, the counts were 
extracted within a cut radius of the FWHM of the PSPC PSF around the HRI
position, and a local background, determined in a concentric ring
from $1\times$ to $1.5\times$ the FWHM around the source position, was 
subtracted. 
For all HRI sources, for which no PSPC source with a 
likelihood $\ge 9$ was found, upper limits (2$\sigma$), have been calculated.

\vfill\pagebreak
\label{masterlistkap}
\setcounter{table}{3}
\unitlength=1cm
\begin{picture}(18,23.5)
\put(0,0){\includegraphics[width=18cm,height=23.5cm,clip=true]{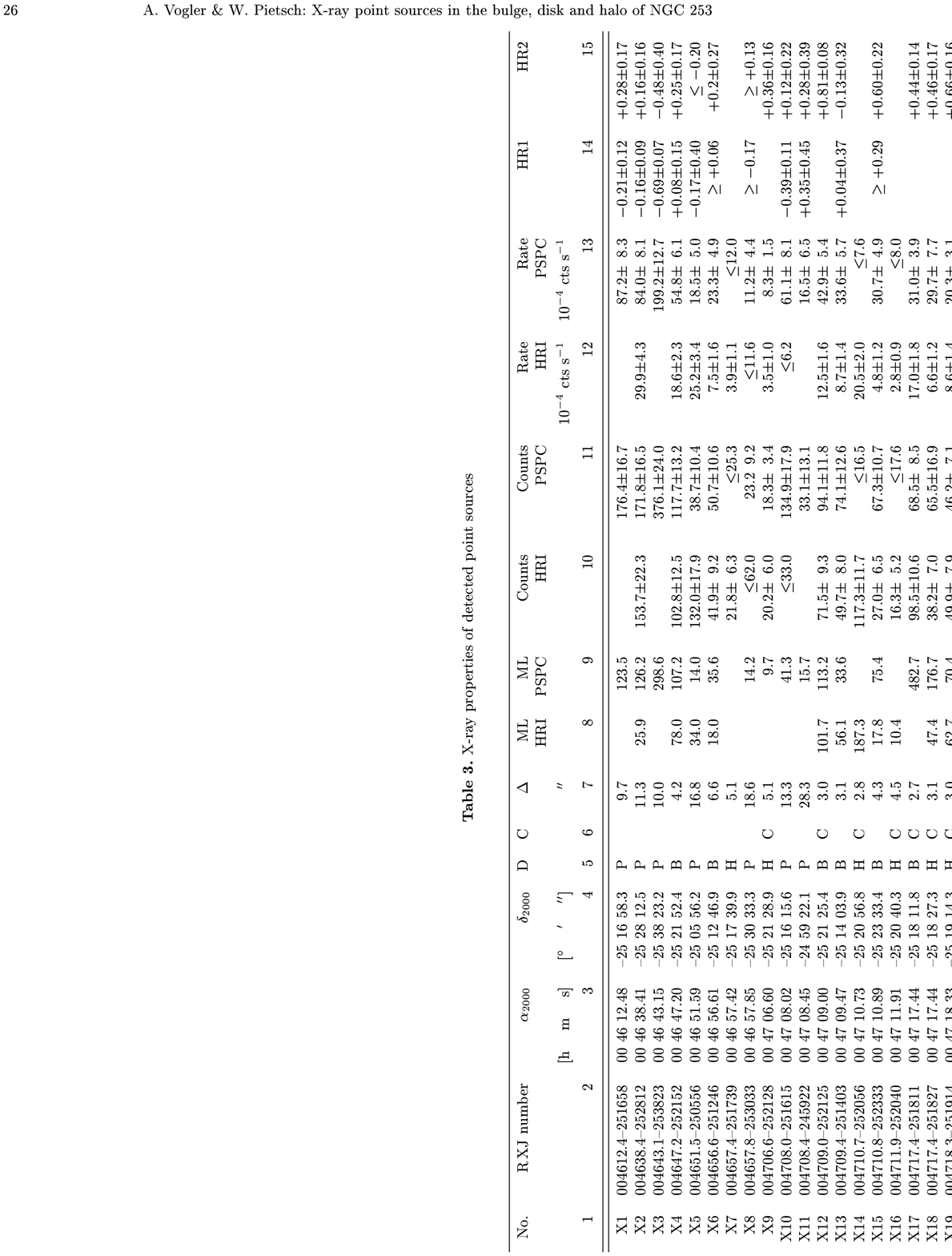}}
\end{picture}\clearpage
\begin{picture}(18,23.5)
\put(0,0){\includegraphics[width=18cm,height=23.5cm,clip=true]{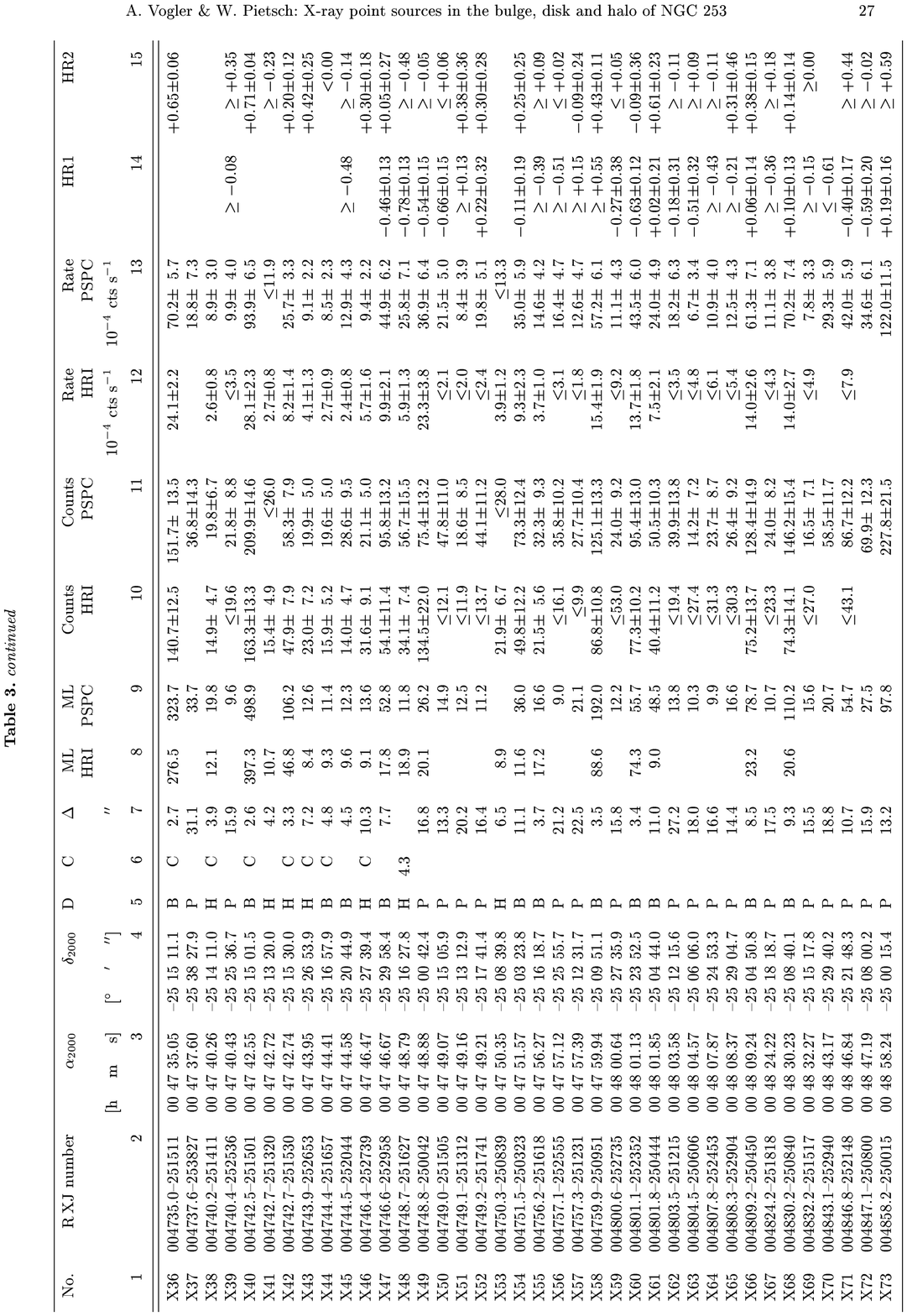}}
\end{picture}\clearpage
\subsection{Time variability investigations} \label{timevardet}
To study time variability of point sources,
counts and count rates were determined for the
individual observation blocks listed in Table~\ref{group}. 
The rates were calculated with fixed source positions. 
The `likelihood ratio test' (cf., 
e.g. Cash 1979, Hogg \& Tanis 1983)
was used to test for variability. Since the comparison of HRI and PSPC count 
rates depends on the assumed spectral model (cf. Table~\ref{ecfs}),
the HRI or PSPC observations were analyzed 
separately.

   \begin{table*}
\caption{Energy conversion factors for the ROSAT HRI and PSPC for different
         spectral models} 
\label{ecfs}
         \begin{flushleft} 
         \begin{tabular}{lccc}
\hline
            \noalign{\smallskip}
 Model & ECF$_{\rm HRI}$$^\star$ & ECF$_{\rm PSPC}$$^\star$
       &$\displaystyle{{\rm ECF_{\rm HRI}\over ECF_{\rm PSPC}}}$\\
            \noalign{\smallskip}
\hline
            \noalign{\smallskip}
 thermal Bremsstrahlung, $T=5~$keV$^\dagger$ & 4.17 & 1.29 & 3.23   \\
 thermal Bremsstrahlung, $T=0.5~$keV & 4.55 & 1.10 & 4.14 \\
 thin thermal plasma, $T=0.3$~keV    &3.77 &1.12 &3.36    \\
            \noalign{\smallskip}
\hline
	 \end{tabular}
	 \end{flushleft}
\vskip-.5cm
\[
\begin{array}{lp{0.95\linewidth}}
^\star    & Energy conversion factors in units 
            of $10^{-11}$~erg~cm$^{-2}$~cts$^{-1}$ \\
^\dagger & Assumed for the conversion from count rates to fluxes
            for point sources\\
\end{array}
\]
   \end{table*}

The `likelihood ratio test' for the six HRI blocks was performed as follows: 
one starts with the hypothesis of a time constant source. 
Count rates $r_{\rm i}$ during
the block i are defined by the ratio of the detected counts $n_{\rm i}$
and the observation time $t_{\rm i}$, and these rates 
should be equal for each block:  
$\bar{r} = \big(\sum_{\rm i=1}^6 {\rm n}_{\rm i}\big) / \big(
\sum_{\rm i=1}^6 t_{\rm i}\big)$, in the case of no variability.
The likelihood ratio is defined as 
$L_{\rm LRT} = 2\times  \sum_{\rm i=1}^6 \,  {\rm n}_{\rm i}
\cdot \ln \big({r_{\rm i}\over \bar{r}}\big)$. If the hypothesis of no 
variability is true, then $L_{\rm LRT}$ due to the counting statistics
has a $\chi^2$ distribution with 5 
degrees of freedom.

\section{Results}
\subsection{HRI and PSPC point sources in the NGC~253 field}
In the NGC~253 field (Figs. 1, 2 and 4) 73 X-ray point sources 
are detected, 23 with both detectors 
(marked in the figures as squares), and 22 and 28 of which exclusively 
with the HRI (hexagons) and PSPC (diamonds), respectively.

Assuming a 5~keV
thermal Bremsstrahlung spectrum (0.1--2.4~keV band and corrected for Galactic
foreground absorption, cf. Table~\ref{ecfs}), the HRI count rates convert to 
fluxes between $8.5\times 10^{-15}$~erg~s~cm$^{-2}$ (source X29) and 
$8.4\times 10^{-13}$~erg~s~cm$^{-2}$ (X34). Similarly,  
PSPC derived fluxes span $8.8\times 10^{-15}$~erg~s~cm$^{-2}$ (X63) to
$9.8\times 10^{-13}$~erg~s~cm$^{-2}$ (X34). As one can see, the longer
observation time for the HRI -- about a factor of three with respect to the
PSPC observation time -- makes up for the lower HRI sensitivity leading to 
almost identical detection limits for the PSPC and HRI observations. 

To allow a crude estimation of the spectral properties of the point sources, 
their hardness ratios (columns 14 and 15 of Table~3) were calculated.
To first order, HR1 traces
the absorption and, to a lesser degree, the hardness of the spectrum, while
HR2 mainly traces the hardness of the spectrum (cf. e.g. Vogler \& Pietsch
1996). The hardness ratios are useful in estimating 
the absorption and spectral 
behavior of point sources, for which the low photon statistics do not
allow spectral investigations. For non-confused sources 
the cut diameters for the sub-bands are chosen 
according to the PSF FWHM for the corresponding off-axis angle
and energy sub-band, and
the background is subtracted with the help of the background
maps. In cases where the error exceeds the counts in one sub-band, only
an upper (lower) limit of the hardness ratio is calculated. For that, the
counts of the non detected band are chosen to be equal to the
upper limit ($2\sigma$). 
Where the errors exceed the counts in both sub-bands, 
no hardness ratios are calculated.
For confused sources, HR1 is not calculated, since 
problems might arise on account of the large
extraction radius used for the PSPC soft band and, to calculate the HR2,
counts
and errors in the hard1 and hard2 band were deduced with the help of the 
multi source fit technique.

\label{masterlistkap}
The light curves of all sources of the catalog (excluding X3 and X37 which
are located at the edge of the calculated PSPC images) are presented in 
Fig.~5. If the source was detected during an observation block 
with a count rate exceeding its error, filled squares are plotted.
Vertical bars represent $1\sigma$ statistical errors.
Upper limits ($2\sigma$) are indicated with open squares in the case
of non-detections. Horizontal lines give the mean count rates during the
whole HRI (blocks 1, 4-8) or PSPC (blocks 2, 3) observations.
Solid lines represent sources that were detected, 
dashed lines represent sources for which
only upper limits could be calculated. In general, the mean count rates 
are in good agreement with the results presented in Table~3. 
The PSPC mean count rates are slightly different from Table~3 for some sources 
located in diffuse emission regions (e.g. X24 or X27).

   \begin{table*}
      \caption{Time variability investigations of the different HRI and PSPC
               observation blocks}
         \label{lightcurves}
         \begin{flushleft}
         \begin{tabular}{rrrrrrr}
            \hline
            \noalign{\smallskip}
No.   & H$^\star$ & P$^\star$   
& Rate$^{\rm max}_{\rm HRI}$ & Rate$^{\rm max}_{\rm PSPC}$ 
             & Flux$^{\rm max}_{\rm HRI}$ & Flux$^{\rm max}_{\rm PSPC}$ \\
            \noalign{\smallskip}
   &$\sigma$ &$\sigma$ &$10^{-3}$~cts~s$^{-1}$ &$10^{-3}$~cts~s$^{-1}$ 
&$10^{-14}$~erg~s$^{-1}$~cm$^{-2}$
&$10^{-14}$~erg~s$^{-1}$~cm$^{-2}$
\\
\noalign{\smallskip} \hline \noalign{\smallskip}
%
% < ******* now row 02 is following ******* >
%
X2 &
$4.1$ &
& &
$4.2$ $\pm$
$
1.1$ &
&
$17.6$ $\pm$
$
4.6$ 
 \\
%
% < ******* now row 05 is following ******* >
%
X5 &
$5.8$ &
&
$1.6$ $\pm$
$
0.8$ &
&
$6.8$ $\pm$
$
3.2$ &
 \\
%
% < ******* now row 07 is following ******* >
%
X7 &
&
$3.6$ &
&
$2.0$ $\pm$
$
0.6$ &
&
$2.6$ $\pm$
$0.8$ \\
%
% < ******* now row 11 is following ******* >
%
X11 &
 &
$3.5$ &
&
$2.8$ $\pm$
$
0.8$ &
&
$3.6$ $\pm$
$1.1$ \\
%
% < ******* now row 12 is following ******* >
%
X12 &
$9.5$ &
$8.9$ &
$3.1$ $\pm$
$
0.5$ &
$6.0$ $\pm$
$
0.2$ &
$12.9$ $\pm$
$
2.0$ &
$7.7$ $\pm$
$0.3$ \\
%
% < ******* now row 13 is following ******* >
%
X13 &
 &
$3.2$ &
 &
$5.9$ $\pm$
$
0.9$ &
 &
$7.5$ $\pm$
$1.2$ \\
%
% < ******* now row 14 is following ******* >
%
X14 &
$11.3$ &
 &
$4.3$ $\pm$
$
0.7$ &
 &
$18.0$ $\pm$
$
2.9$ &
 \\
%
% < ******* now row 15 is following ******* >
%
X15 &
$3.0$ &
&
$0.9$ $\pm$
$
0.3$ &
 &
$3.6$ $\pm$
$
1.2$ &
 \\
%
% < ******* now row 17 is following ******* >
%
X17 &
$6.4$ &
&
$4.2$ $\pm$
$
0.7$ &
&
$17.5$ $\pm$
$
2.9$ & \\
%
% < ******* now row 20 is following ******* >
%
X20 &
$4.6$ &
&
$1.6$ $\pm$
$
0.5$ &
 &
$6.6$ $\pm$
$
2.2$ &
 \\
%
% < ******* now row 22 is following ******* >
%
X22 &
$3.6$ &
 &
$2.3$ $\pm$
$
0.5$ &
 &
$9.5$ $\pm$
$
2.3$ &
 \\
%
% < ******* now row 28 is following ******* >
%
X28 &
$5.9$ &
 &
$1.8$ $\pm$
$
0.5$ &
 &
$7.5$ $\pm$
$
2.0$ &
 \\
%
% < ******* now row 30 is following ******* >
%
X30 &
$4.7$ &
 &
$2.6$ $\pm$
$
0.8$ &
 &
$11.0$ $\pm$
$
3.3$ &
 \\
%
% < ******* now row 33 is following ******* >
%
X33 &
 &
$8.3$ & &
$45.0$ $\pm$
$
1.5$ &
&
$58.5$ $\pm$
$2.0$ \\
%
% < ******* now row 35 is following ******* >
%
X35 &
$4.5$ &
 &
$1.3$ $\pm$
$
0.4$ &
 &
$5.4$ $\pm$
$
1.8$ &
 \\
%
% < ******* now row 40 is following ******* >
%
X40 &
$3.7$ &
$3.8$ &
$4.5$ $\pm$
$
0.7$ & $^\dagger$
%$9.1$ $\pm$
%$
%0.1$ 
&
$18.8$ $\pm$
$
2.9$ & $^\dagger$
%$11.8$ $\pm$
%$0.2$ 
\\
%
% < ******* now row 46 is following ******* >
%
X46 &
$5.4$ &
 &
$4.4$ $\pm$
$
1.6$ & &
$18.5$ $\pm$
$
6.6$ &
\\
%
% < ******* now row 54 is following ******* >
%
X54 &
$4.3$ &
 &
$2.1$ $\pm$
$
1.3$ &
 &
$8.9$ $\pm$
$
5.3$ &
 \\
%
% < ******* now row 57 is following ******* >
%
%X57 &
% &
%$3.5$ &
%&
%$1.9$ $\pm$
%$
%0.6$ &
% &
%$2.5$ $\pm$
%$0.8$ \\
%
% < ******* now row 61 is following ******* >
%
X61 &
$4.6$ &
 &
$1.8$ $\pm$
$
0.5$ &
 &
$7.3$ $\pm$
$
2.2$ &
 \\
%
% < ******* now row 64 is following ******* >
%
X64 &
 &
$2.9$ &
 &
$2.0$ $\pm$
$
0.7$ &
 &
$2.6$ $\pm$
$0.9$ \\
%
% < ******* now row 66 is following ******* >
%
X66 &
$3.7$ &
 &
$2.9$ $\pm$
$
0.8$ &
 &
$12.3$ $\pm$
$
3.2$ & \\
%
% < ******* now row 68 is following ******* >
%
X68 &
$5.1$ &
 &
$3.5$ $\pm$
$
0.8$ &
 &
$14.5$ $\pm$
$
3.3$ &
\\
%
% < ******* now row 70 is following ******* >
%
X70 &
 &
$6.2$ &
&
$5.4$ $\pm$
$
1.0$ &
 &
$7.0$ $\pm$
$1.3$ \\
%
% < ******* now row 71 is following ******* >
%
X71 &
 &
$3.3$ &
 &
$5.8$ $\pm$
$
0.9$ &
 &
$7.5$ $\pm$
$1.2$ \\
            \noalign{\smallskip}
            \hline
	 \end{tabular}
	 \end{flushleft}
\vskip-.5cm
\[
\begin{array}{lp{0.95\linewidth}}
$$^\star$$ & Gaussian significance for the detection of time variability during 
            the individual HRI (H) or PSPC (P) observation blocks according to 
            Sect.~\ref{timevardet}\\
^\dagger  & Time Variability during the PSPC blocks established. 
            However, the maximum count rate (flux) 
            is in the order of the integral count rate (flux)
            of the total observation. The problem might be due to diffuse
            emission surrounding the source as well as due to the very nearby 
            source X42 \\
\end{array}
\]
   \end{table*}

To further characterize the point sources of the catalog, they were 
checked for time variability using the likelihood ratio test. 
The likelihood ratios
of the time variability test performed for the HRI and PSPC
observations (Sect.~\ref{timevardet})
can be transformed into a probability 
that the source is time-variable (cf., e.g. Bronstein 1985). 
In Table~\ref{lightcurves} we summarize the information on those 24 sources of 
the total of 73, for which
a Gaussian significance for time variability with $\sigma \gid 3$ was found
within the HRI or PSPC observation blocks. Besides the significances,
maximum count rates and fluxes (cf. Table~\ref{ecfs} 
for the conversion factor and assumed model) are given.

{
\begin{figure*}
\includegraphics[bb=53 81 479 759,width=18cm,height=20cm,clip=true]{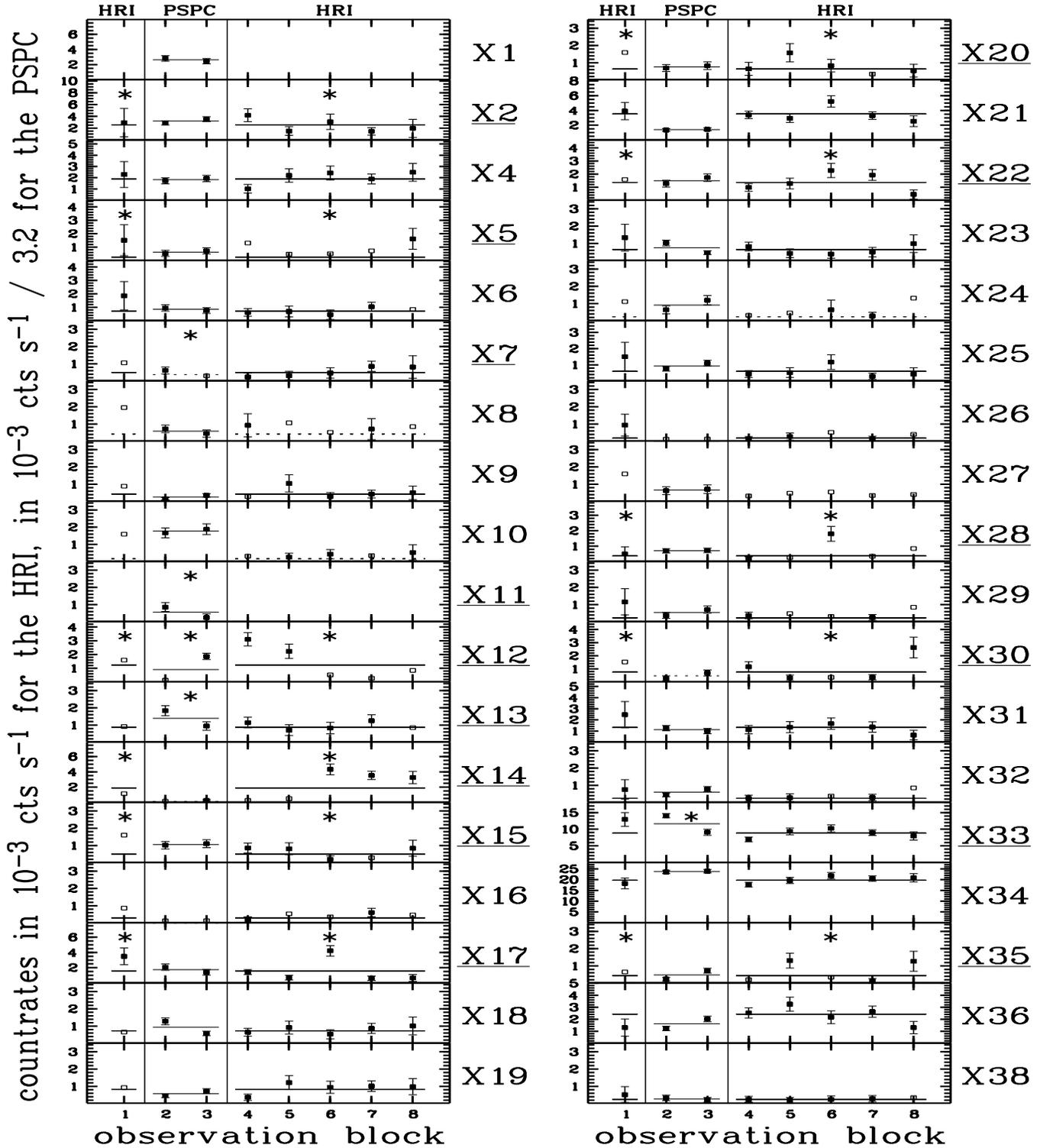}
\caption{
Light curves of X-ray sources in the NGC~253 field (excluding X3 and X37 which
were located at the edge of the investigated field). 
Where the source was detected during an observation block 
with a count rate exceeding the error filled squares are used as symbols and
the errors are indicated as bars. In the case of errors exceeding the count 
rates upper limits ($2\sigma$) are plotted as open squares. 
The horizontal lines give the
mean count rate calculated as explained Sect.~\ref{timevardet}. 
Solid lines represent sources
that were detected according to Table~3, dashed lines represent 
sources for which only upper limits could be calculated. 
Stars above the HRI or PSPC light curves mark in which detector the 
variability was detected. The PSPC count rates have been divided by a factor
of 3.2 before they were plotted in the diagram (this is the ratio between
the energy conversion factors of the HRI and PSPC for a 5~keV thermal 
Bremsstrahlung spectrum corrected for Galactic absorption, cf. 
Table~\ref{ecfs}). If one
detects an isolated point source and the spectral model is right, 
the light curve should be a straight line
        }
\end{figure*}
\addtocounter{figure}{-1}
\begin{figure*}
\includegraphics[bb=53 81 479 759,width=18cm,height=20cm,clip=true]{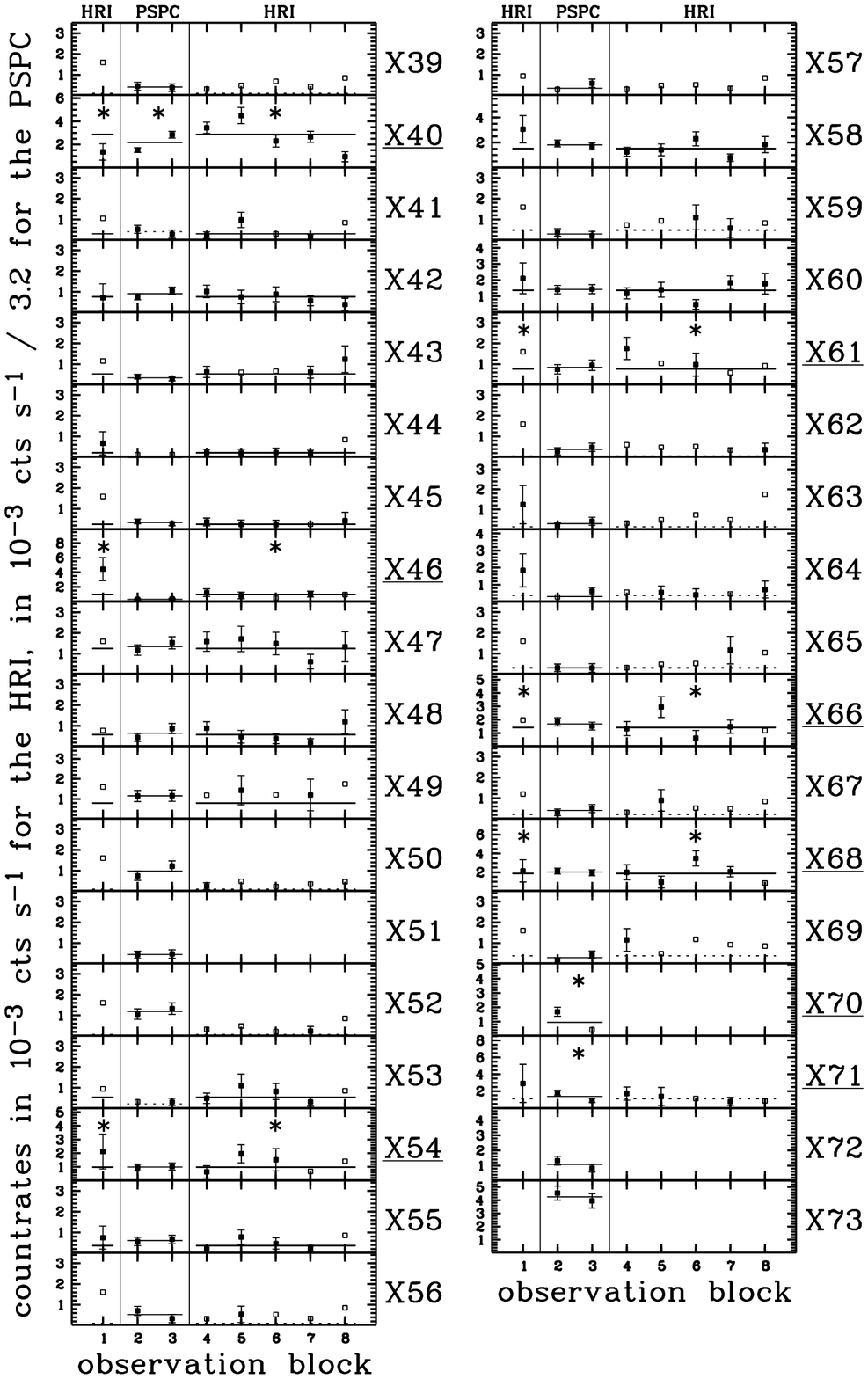}
\caption{\it continued}
\end{figure*}
}

Time variability for a source might also be established via a comparison of
the HRI and PSPC derived fluxes. This method, however, might feign time 
variability if the wrong spectral model is used for count rate to flux 
conversion. We calculated the ratio
of the energy conversion factors for our assumed model (5~keV thermal
Bremsstrahlung), a 0.5~keV thermal Bremsstrahlung spectrum and a thin
thermal plasma of 0.3~keV
(cf. Table~\ref{ecfs}). The examples indicate that a wrong spectral model
might fake time variability to the order of up to 30$\%$ or even more for
extreme examples, e.g. a plasma with a temperature below 0.2~keV. 
Keeping this reservation in mind, 
time variability is suggested with $\sigma\gid 3$ for four
additional sources which were not already 
picked up by investigating HRI or PSPC blocks, individually: 
X10, X21, X24, X34. 
In the case of X10 and X24, the HRI did not pick up a point-like source. 
The sources X21 and X34 are located within the diffuse emission of 
the NGC~253 bulge and disk and it can not be excluded that the 
PSPC determined counts are affected by subtracting a wrong background.

\subsection{Resolving the complex nuclear emission area}
\label{verybright}
Two bright sources, X33 and X34, are detected 
in the central region of NGC~253, both embedded in a complicated diffuse
X-ray emission structure visible in the PSPC and HRI images (Figs.~1 to 4).
We tried to disentangle the emission components with the help of the spatial
resolution of the HRI and the spectral resolution of the PSPC.

The luminosities as suggested by the HRI count rates are 
$(3.0$$\pm$$0.1)\times 10^{38}$~erg~s$^{-1}$ and 
$(6.6$$\pm$$0.2)\times 10^{38}$~erg~s$^{-1}$ for X33 and X34, respectively.
The position of the source X34 coincides with the optical center of NGC~253, 
X33 being located $\sim 20''$ to the south. 
Together with the sources X33 and X34,
the counts within the central emission region
encircled by the lowest contour in Fig.~3 amount to 
$2803\pm53$ HRI counts.
We focus in this paper on the
point sources and will discuss the diffuse emission components in PEA.

\subsubsection{Spatial analysis of the HRI data}
\label{latenighttrain}
The HRI detection algorithm flagged
X33 and X34 as extended (FWHM of extent $8''$ and $11''$, respectively).
X33 and X34 are, however,  
embedded in a region of diffuse emission and the detection algorithm may
be fooled if this background is not modeled correctly by the background map.
To investigate if these sources really are point sources inside diffuse
emission, radial surface
brightness profiles from 0 to 15$''$ radius were calculated. 
To visualize the investigated region, 
circles with radii of 15$''$ around the sources are sketched in Fig.~3. 
The surface brightness profile
of X34 is centered on the intensity maximum correlated with X34. Due to the
extent of the source and the slower decay of the intensity towards the
east than to the west (cf. Fig.~3), the source position determined by the
maximum likelihood algorithm and given in Table 3 is located $\sim3\arcsec$
southeast from this maximum. 
The width of the individual rings of the profiles is
3$''$. In principal, one could compare the
profiles of X33 and X34 to analytical models of the PSF. However, as there
was no point source in the HRI field of view,
bright enough to allow the correction of the pointing
positions on time intervals shorter than the wobble period ($\sim$400 s), 
the SASS attitude solution could only be improved on longer time scales 
(see Sect. 2.1), and one has to expect that the PSF is 
slightly broadened due to residual artifacts of the satellite
wobble movement. We therefore 
choose to compare the surface brightness profiles 
of X33 and X34 to the profiles of the bright unconfused point-like
sources X21 and X36, which were collected with
the same attitude solution as X33 and X34 (Fig.~6). 
Within the off-axis angles of X21
and X36 (4\farcm6 and 2\farcm 2, respectively), the HRI PSF is not expected
to deteriorate compared to the central sources. 
\label{dieguteanmerkung}

\begin{figure}
\includegraphics[bb=60 129 321 496,width=8.8cm,clip=true]{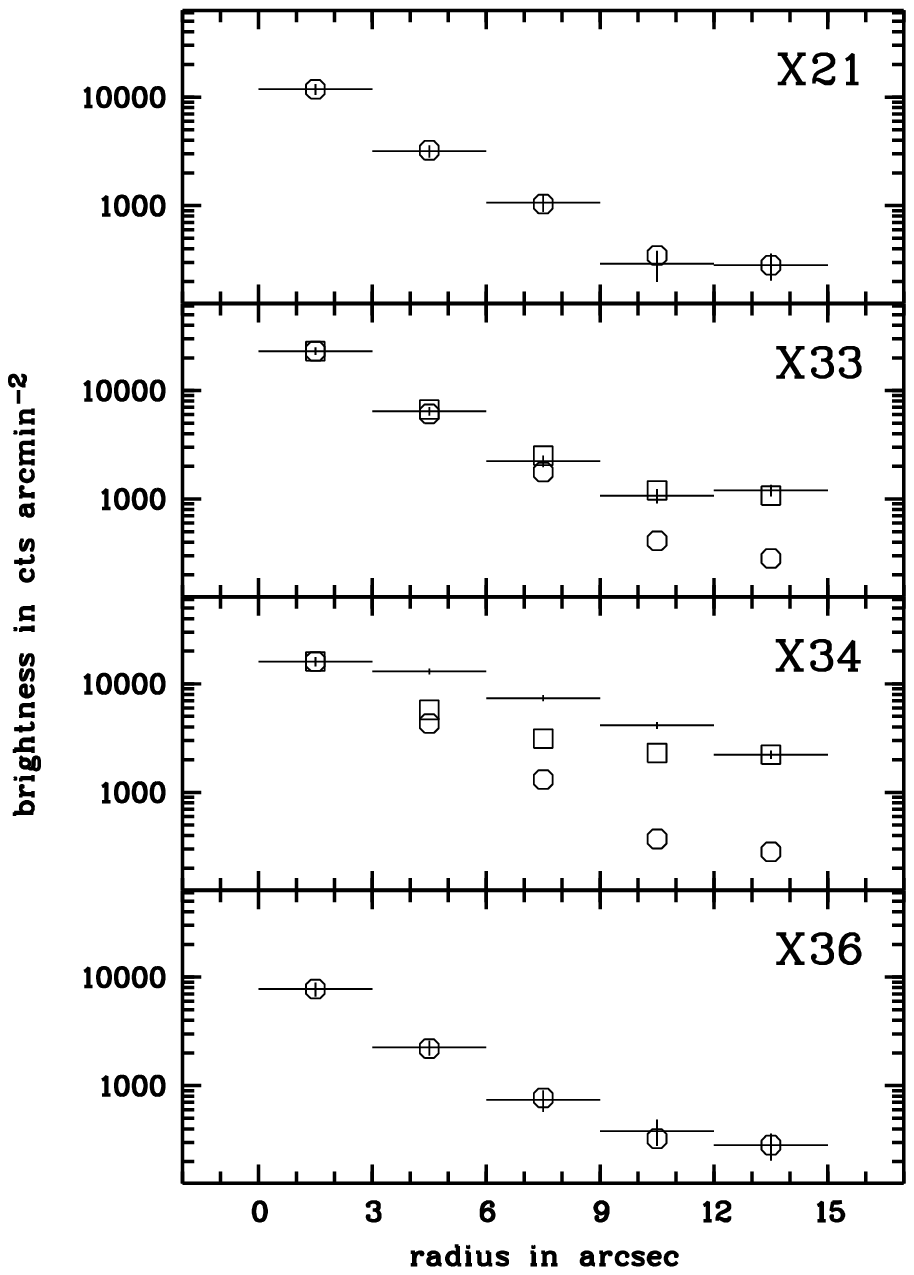}
\caption{ %{\bf \ \ X21 tiefer \ \ }
Radial surface brightness profiles for the sources X33, X34, X21 and X36, 
calculated with a radial binwidth of
3$''$ from the HRI observations. The crosses mark the measurements for the
individual sources, the length along the x and y axis indicate the binwidth
and the error of the brightness, respectively. The open circles show the
`experimental' PSF model as obtained from the point-like
sources X21 and X36. The PSF predictions for point-like sources at the 
position of X33 and X34 (cf. Sect~\ref{latenighttrain}) have been marked as
open squares
        }
\label{radial}
\end{figure}

The `experimental' PSF is calculated by averaging the surface 
brightness profiles of X21 and X36. The maximum of the PSF is normalized to the
inner radial bin ($0'' - 3''$). As can be seen in Fig. 6 (circles plotted
over the profiles), this PSF model represents a good description 
of the brightness profiles of X21 and X36. 
To adapt the model for X33 and X34, the maximum was again normalized to the
inner radial bin. The fact that X33 and X34 are sources embedded 
in a diffuse emission region was taken into account by normalizing 
the background in the 
outermost radial bin ($12'' - 15''$).  As can be seen in Fig. 6 (squares 
plotted over the profiles), this model is a good approximation to the 
profile of X33, though it cannot describe the slow decay of the surface
brightness of X34. 
X33 therefore is likely to be a point source.  Integrating the PSF model,
359$\pm$21 counts ($f_{\rm x}=2.5\times 10^{-13}
$~erg~s$^{-1}$~cm$^{-2}$, $L_{\rm x}=2.0\times 10^{38}$~erg~s$^{-1}$)
are deduced for X33, a number significantly 
lower than the number of counts determined by the detection algorithm
(520$\pm$24 counts). That number has been overestimated as 
X33 was fitted as an extended source
and not the whole diffuse background was taken into account in the background 
map.
On the contrary, the X-ray emission of X34
cannot be due to a single point-like source. Within a circle of 9$''$
around X34, 703$\pm$26 counts were measured after subtraction of the
background, whereas the PSF model 
predicts less than 300 counts for a point source.

\subsubsection{Spectral analysis of the PSPC data}

\begin{figure}
\unitlength=1cm
\begin{picture}(8.8,11.1)
\put(0,0){\includegraphics[bb=53 127 394 268,width=8.8cm,clip=true]{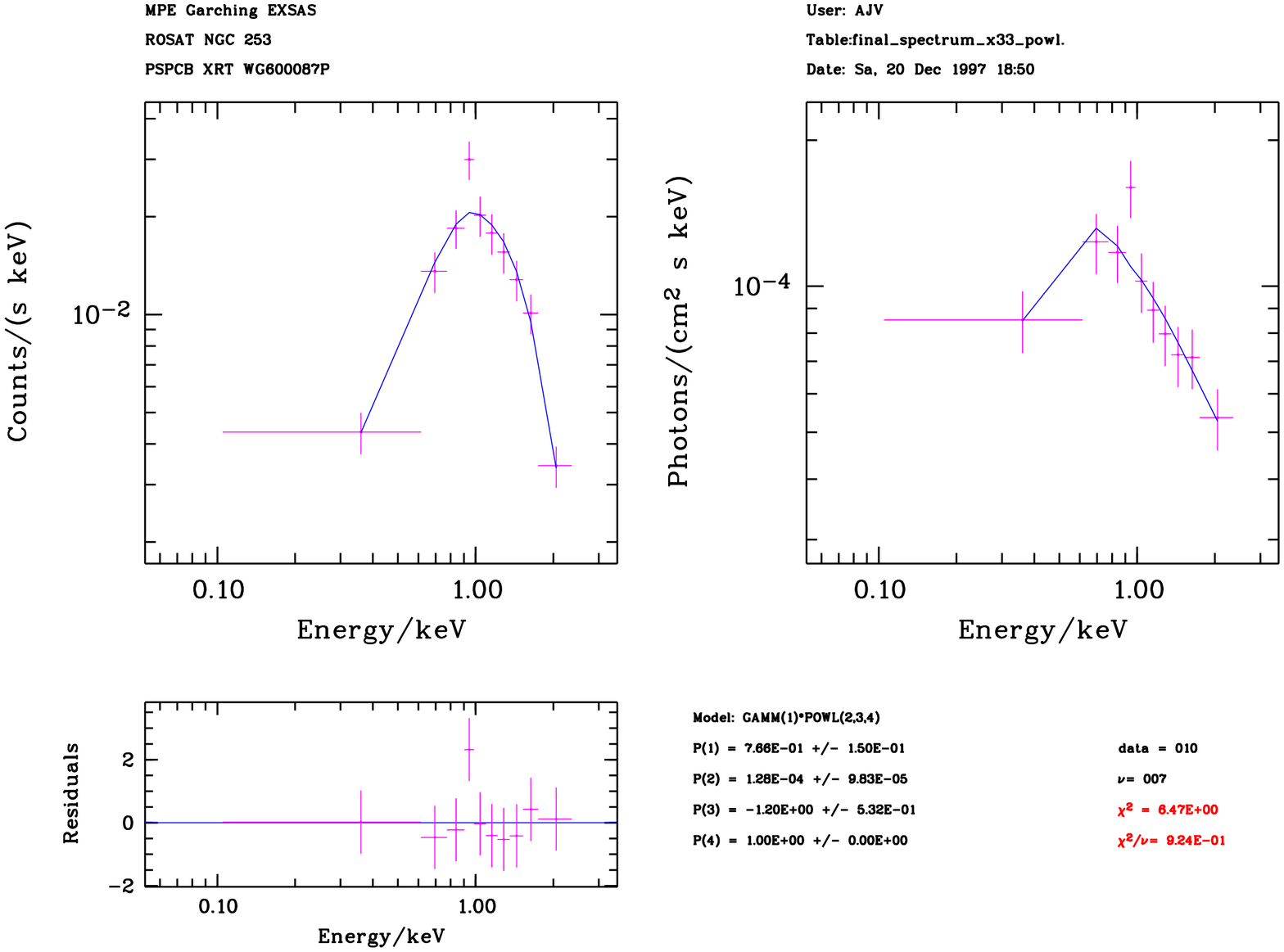}}
\put(0,4){\includegraphics[bb=416 320 757 599,width=8.8cm,clip=true]{powl_x33.ps}}
\put(2.6,10){
\parbox{5cm}{\large Power law fit \\ of X33}
            }
\put(2.6,6.2){
\parbox{5cm}{Fit parameters:
            }
           }
\put(2.6,5.7){$\phantom{00}N_{\rm H}=6.5^{+0.9}_{-0.3}\times10^{20}$~cm$^{-2}$
\parbox{5cm}{
            }
           }
\put(2.6,5.2){
\parbox{5cm}{$\phantom{007}\Gamma=1.2^{+0.6}_{-0.2}$
            }
           }
\end{picture}
\caption{
Results of a power law fit to X33. {\it Top panel:}
Flux of the observed X-ray emission
normalized to the energy, in photons/($\rm cm^{-2}~s^{-1}~keV^{-1}$), the
crosses represent the observed flux (count rates defolded by the spectral
model), the solid curve gives the best fit. 
{\it Bottom panel:} residuals of the fit
        }
\label{fitplot}
\end{figure}

To further investigate the structure of the point-like source X33, we made use
of the spectral capabilities of the PSPC. A source spectrum was extracted 
with a cut diameter of $30''$. The very small extraction diameter
and the varying PSF of the individual channels were corrected for using standard
EXSAS procedures. Background was subtracted from a 
source-free region outside the disk and halo of the galaxy, leading to 
a raw spectrum of X33 containing 352$\pm$23 counts.
This raw spectrum was binned into energy bands to give 
a signal to noise ratio $\gid7$.
Simple spectral models, a power law (POWL), thermal
Bremsstrahlung (THBR) and a thin thermal plasma (THPL) model were fitted.
With free absorption,
free normalization and free index/temperature, our fits had seven degrees of
freedom. Formally, all models achieved $\chi^2/\nu$ values between 0.9 and 1.0.
However, as can be seen from the suggested temperatures of the thermal models 
(THBR: $T>100$~keV; 
THPL: $T=17$~keV), they fall in a range that cannot be
properly constrained by ROSAT. 
The POWL fit (Fig.~7) resulted in an intrinsic
absorption (after subtracting the Galactic
foreground) of $N_{\rm H}=6.5^{+0.9}_{-0.3} \times 10^{20}$~cm$^{-2}$,
a photon index of $1.2_{-0.3}^{+0.6}$ and flux of 
$3.5^{+0.6}_{-0.5}\times 10^{-13}$~erg~s$^{-1}$~cm$^{-2}$ (0.1--2.4 keV). 
The errors (1$\sigma$) for the fit parameters were
calculated with the help of the error ellipses. The residuals 
of the fit are very small with the exception of the bin around 0.95~keV.
Spectra of X34 (cf. PEA) indicate the presence
of a thermal emission component with a temperature around 1~keV. 
Assuming that this 
diffuse emission is also contributing at the position of X33, 
the high residual of
the bin around 0.95~keV could be explained. The photon index of X33 suggests
a very hard intrinsic spectrum. In comparison, the index
of a POWL fit for X34 (same extraction radius chosen as for X33) is 
$3.0_{-0.7}^{+0.8}$ and suggests a much softer spectrum. 
The absorption of X33 ($N_{\rm H}=6.5 \times 10^{20}$~cm$^{-2}$) is clearly 
less than the absorption of X34 ($N_{\rm H}=(2-3)\times 10^{21}$~cm$^{-2}$). 
The spectrum of X34 together with the spectrum of the underlying diffuse
emission is discussed in detail in PEA.

Combining the PSPC spectrum with the results of the surface brightness
profiles obtained from the HRI observations, one derives an improved
luminosity for X33 of $L_{\rm x} =
4.0\times 10^{38}$~erg~s$^{-1}$, corrected for the Galactic foreground and
absorption within NGC~253. This luminosity is significantly lower than the 
one deduced from our detection catalog and time variability investigations. 

\subsection{Point sources within the NGC~253 disk}
\label{disksource}
Of the 73 field sources, 32 are located within the $D_{25}$ ellipse of NGC~253
and attributed to the NGC~253 disk. Some of these sources may be spurious
detections caused by diffuse filamentary X-ray emission features which 
are most clearly seen in the ROSAT PSPC image (Fig.~4), covering 
the disk and halo of NGC~253. For the HRI, 
similar problems exist in the 
central emission region shown in Fig.~3 and in the inner spiral arms of NGC~253 
(indeed, one source has already been removed when creating the HRI
source catalog). For the PSPC, we exclude from the further discussion 
those sources in the NGC~253 disk, which are located in regions 
of diffuse X-ray emission {\it and} which have only been detected with the 
PSPC, namely X50, X51, X52, X57 and X62. A visual check and a comparison
with our list of transients verifies that no PSPC bright sources are rejected 
by this procedure.

All 27 remaining sources have been detected with the HRI, the count rates 
lying between $2.0\times 10^{-4}$~cts~s$^{-1}$ (X26) and 
$2.0\times 10^{-2}$~cts~s$^{-1}$ (X34). At the distance of NGC~253, this 
converts to luminosities 
between $6.7\times 10^{36}$~erg~s$^{-1}$ 
and $6.6\times 10^{38}$~erg~s$^{-1}$. 23 of these 
sources have been also detected with 
the PSPC, with count rates from $8.3\times 10^{-4}$~cts~s$^{-1}$ (X9)
to $7.6\times 10^{-2}$~cts~s$^{-1}$ (X34), converting to luminosities
between $8.5\times 10^{36}$~erg~s$^{-1}$ and $7.9\times 10^{38}$~erg~s$^{-1}$.

\begin{table}
\caption{Luminosities of sources located within the area covered by 
         the $D_{25}$ ellipse of NGC~253}
\label{luminosities}
\begin{flushleft}
\begin{tabular}{rrrrr}
\hline
\noalign{\smallskip}
No.&
$L_{\rm x}^{\rm HRI}~^\star$ &
$L_{\rm x}^{\rm PSPC}~^\star$&
$L_{\rm x,~max}^{\rm HRI}~^\star$ &
$L_{\rm x,~max}^{\rm PSPC}~^\star$ \\
\noalign{\smallskip}
\hline
\noalign{\smallskip}
%
% < ******* now row 09 is following ******* >
%
X9 &
$1.2$ $\pm$
$
0.3$ &
$0.8$ $\pm$
$
0.2$ &
$
$
 &
$
$ \\
%
% < ******* now row 12 is following ******* >
%
X12 &
$4.2$ $\pm$
$
0.5$ &
$4.5$ $\pm$ $0.6$ &
$10.3$ $\pm$
$
1.6$ &
$6.1$ $\pm$
$0.2$ \\
%
% < ******* now row 14 is following ******* >
%
X14 &
$6.8$ $\pm$
$
0.7$ &
$\le$ $0.8$
 &
$14.3$ $\pm$
$
2.3$ &
$
$ \\
%
% < ******* now row 15 is following ******* >
%
X15 &
$1.6$ $\pm$
$
0.4$ &
$3.2$ $\pm$
$
0.5$ &
$2.9$ $\pm$
$
1.0$ &
$
$ \\
%
% < ******* now row 16 is following ******* >
%
X16 &
$0.9$ $\pm$
$
0.3$ &
$\le$ $0.8$
 &
$
$
 &
$
$ \\
%
% < ******* now row 17 is following ******* >
%
X17 &
$5.6$ $\pm$
$
0.6$ &
$3.2$ $\pm$ $0.4$
 &
$14.0$ $\pm$
$
2.3$ &
$
$ \\
%
% < ******* now row 18 is following ******* >
%
X18 & $2.2$ $\pm$ $0.4$ & $ 3.0$ $\pm$ $0.8 $ \\
%
% < ******* now row 19 is following ******* >
%
X19 &
$2.9$ $\pm$
$
0.5$ &
$2.1$ $\pm$
$
0.2$ &
$
$
 &
$
$ \\
%
% < ******* now row 21 is following ******* >
%
X21 &
$11.2$ $\pm$
$
0.8$ &
$7.3$ $\pm$
$
0.6$ &
$
$
 &
$
$ \\
%
% < ******* now row 23 is following ******* >
%
X23 &
$2.3$ $\pm$
$
0.4$ &
$2.5$ $\pm$
$
0.4$ &
$
$
 &
$
$ \\
%
% < ******* now row 25 is following ******* >
%
X25 &
$2.3$ $\pm$
$
0.5$ &
$2.4$ $\pm$
$
0.2$ &
$
$
 &
$
$ \\
%
% < ******* now row 26 is following ******* >
%
X26 &
$0.7$ $\pm$
$
0.2$ &
$\le$ $1.4$
 &
$
$
 &
$
$ \\
%
% < ******* now row 28 is following ******* >
%
X28 &
$1.8$ $\pm$
$
0.4$ &
$1.2$ $\pm$ $0.2$
 &
$6.0$ $\pm$
$
1.6$ &
$
$ \\
%
% < ******* now row 29 is following ******* >
%
X29 &
$0.8$ $\pm$
$
0.3$ &
$1.8$ $\pm$
$
0.5$ &
$
$
 &
$
$ \\
%
% < ******* now row 32 is following ******* >
%
X32 &
$1.1$ $\pm$
$
0.3$ &
$1.2$ $\pm$ $0.2$
 &
$
$
 &
$
$ \\
%
% < ******* now row 33 is following ******* >
%
X33 &
$29.5$ $\pm$
$
1.4$ &
$26.4$ $\pm$ $1.1$ &
$
$
 &
$30.2$ $\pm$
$2.0$ \\
%
% < ******* now row 34 is following ******* >
%
X34 &
$66.2$ $\pm$
$
2.0$ &
$77.9$ $\pm$
$
1.8$ &
$
$
 &
$
$ \\
%
% < ******* now row 35 is following ******* >
%
X35 &
$2.1$ $\pm$
$
0.5$ &
$1.2$ $\pm$ $0.3$
 &
$4.3$ $\pm$
$
1.4$ &
$
$ \\
%
% < ******* now row 36 is following ******* >
%
X36 &
$8.0$ $\pm$
$
0.7$ &
$7.2$ $\pm$
$
0.6$ &
$
$
 &
$
$ \\
%
% < ******* now row 38 is following ******* >
%
X38 &
$0.9$ $\pm$
$
0.3$ &
$0.9$ $\pm$
$
0.3$ &
$
$
 &
$
$ \\
%
% < ******* now row 40 is following ******* >
%
X40 &
$9.3$ $\pm$
$
0.8$ &
$9.7$ $\pm$
$
0.7$ &
$14.9$ $\pm$
$
2.3$ &
$^\dagger$ \\
%
% < ******* now row 41 is following ******* >
%
X41 &
$0.9$ $\pm$
$
0.3$ &
$\le$ $1.2$
 &
$
$
 &
$
$ \\
%
% < ******* now row 42 is following ******* >
%
X42 &
$2.7$ $\pm$
$
0.5$ &
$2.6$ $\pm$
$
0.6$ &
$
$
 &
$
$ \\
%
% < ******* now row 44 is following ******* >
%
X44 &
$0.9$ $\pm$
$
0.3$ &
$0.9$ $\pm$
$
0.2$ &
$
$
 &
$
$ \\
%
% < ******* now row 48 is following ******* >
%
X48 &
$1.9$ $\pm$
$
0.4$ &
$2.7$ $\pm$
$
0.7$ &
$
$
 &
$
$ \\
%
% < ******* now row 50 is following ******* >
%
$^\clubsuit$X50 &
$\le$ $0.7$
 &
$2.2$ $\pm$
$
0.5$ &
$
$
 &
$
$ \\
%
% < ******* now row 51 is following ******* >
%
$^\clubsuit$X51 &
$\le$ $0.2$
 &
$0.9$ $\pm$
$
0.4$ &
$
$
 &
$
$ \\
%
% < ******* now row 52 is following ******* >
%
$^\clubsuit$X52 &
$\le$ $0.8$
 &
$2.0$ $\pm$
$
0.5$ &
$
$
 &
$
$ \\
%
% < ******* now row 55 is following ******* >
%
X55 &
$1.2$ $\pm$
$
0.3$ &
$1.5$ $\pm$
$
0.4$ &
$
$
 &
$
$ \\
%
% < ******* now row 57 is following ******* >
%
$^\clubsuit$X57 &
$\le$ $0.6$
 &
$1.3$ $\pm$
$
0.5$ &
$
$ 
% &
%$2.0$ $\pm$
%$0.7$ 
\\
%
% < ******* now row 58 is following ******* >
%
$^\heartsuit$X58 & (5.1) & (5.9) \\ 
%
% < ******* now row 62 is following ******* >
%
$^\clubsuit$X62 &
$\le$ $1.2$
 &
$1.9$ $\pm$
$
0.6$ &
$
$
 &
$
$ \\
\noalign{\smallskip}
\hline
\end{tabular}
\end{flushleft}
\vskip-.5cm
\[
\begin{array}{lp{0.95\linewidth}}
^\star   & Luminosities at the distance of NGC~253 in 
           units of $10^{37}$~erg~s$^{-1}$, cf. Table~5 for assumed model
             \\
^\dagger & Time Variability during the PSPC blocks established. 
           However, the maximum flux is of the order of the integral
           flux of the total observation. The problem might be due to diffuse
           emission surrounding the source as well as due to the very nearby 
           source X42\\
^\clubsuit & Propably due to enhancements of diffuse X-ray emission\\
^\heartsuit    & Background quasar\\
\end{array}
\]
\end{table}

The individual source luminosities are listed in Table~\ref{luminosities}. 
For 8 of the 27 NGC~253 sources, either the ROSAT HRI or PSPC data alone proof 
variability. For the brightest source (X33) statistics were sufficient to
search for time variability on shorter time scales.
Single observation blocks were analyzed, but no short term time 
variability could be established.
Two further time-variable sources
(X21 and X34) are suggested by comparing 
the HRI and PSPC fluxes. The apparent time variability of X34, however, can be 
explained as a side effect of the limited resolution of the PSPC 
hampering the flux determination for this extended source. 
Therefore, excluding X34, the ROSAT results indicate time variability 
for 9 of the 27 sources in the disk of NGC~253. 

Transients form a 
special subclass of the time variable sources. For the purpose of this paper
a source will be called transient if it remains undetected during at
least one observation interval 
(2$\sigma$ detection limits of the individual intervals 
$\sim 1\times 10^{37}$~erg~s$^{-1}$) and shows a monotonic increase/decrease
of the peak luminosity, the
simplest case being a source only detected during one observation block.
This definition of a transient may actually be fulfilled if a source
shows an outburst by a factor of a few because of the limited sensitivity of the
NGC~253 observations, and in this case the definition is less stringent than 
for transients in the Galaxy or the Magellanic clouds.
From their ROSAT light curves X12, X14 are transients in this restricted sence. 

Our investigations establish time variability for all NGC~253 disk sources with 
luminosity maxima above $3\times 10^{37}$~erg~s$^{-1}$, with the
only exception of  
X36, for which no time variability could be found. 
For fainter sources with no established
variability, it remains unclear
whether these sources are time-constant or whether the statistics are too 
low to establish variability.

\begin{table*}
\caption{Results of thermal Bremsstrahlung fits to 
         X12, X17, X21, X36 and X40}
\label{fits}
\begin{flushleft}
\begin{tabular}{lcccccc}
\hline
\noalign{\smallskip} 
No & Counts in 
 &$N_{\rm H}~^\star$ &$T$ & $\chi^2/\,$DOF & $L_{\rm x}^{\rm abs}~^\dagger$ 
                                    & $L_{\rm x}^{\rm unabs}~^\ddagger$\\
 &raw spectrum \\
&cts &$10^{21}~$cm$^{-2}$ & keV & &$10^{38}~$erg~s$^{-1}$  
 &$10^{38}~$erg~s$^{-1}$\\
\noalign{\smallskip}
\hline
\noalign{\smallskip}
X12 & $\phantom{0}89.0\pm13.5$
    & $11^{+\infty}_{-9}$ & $0.45^{+\infty}_{-0.3}$ & $\phantom{0}$8.8\,/\,10
    & 1.2 & 34.5 \\
\noalign{\smallskip}
X17 (+ nearby source X18)& $202.2\pm18.3$
    & $\phantom{0}6.9^{+3.6}_{-3.4}$ & $0.38^{+0.6}_{-0.1}$ & 11.0\,/\,11 
    & 1.6 &27.8 \\
\noalign{\smallskip}
X21 & $196.6\pm19.5$
    & $\phantom{0}1.6^{+3.2}_{-0.9}$ 
    & $1.2^{+16}_{-0.7}$ & 23.5\,/\,21 & 1.1 & $\phantom{0}$2.4 \\
\noalign{\smallskip}
X36 & $195.7\pm 18.3$
    & $\phantom{0}9.5^{+3.5}_{-5.5}$ & $0.38^{+1.1}_{-0.1}$ & 11.0\,/\,21 
    & 0.9 & 37.6 \\
\noalign{\smallskip}
X40 (+ nearby source X42)& $236.1\pm21.1$
    & $\phantom{0}7.0^{+4.0}_{-3.7}$ 
    & $0.40^{+0.7}_{-0.1}$ & $\phantom{0}$9.1\,/\,16 
    & 1.7 & 27.9\\
\noalign{\smallskip}
\hline
\end{tabular}
\end{flushleft}
\vskip-.5cm
\[
\begin{array}{lp{0.95\linewidth}}
^\star    & In excess of the Galactic foreground\\
^\dagger  &  Luminosity corrected for Galactic foreground $N_{\rm H}$. 
             The spectral model as received from the PSPC fit was folded with
             the HRI count rates to obtain the luminosities. For the time
             variable sources X12, X17 and X40 we used the maximum HRI count 
             rates, for X21 and X36 we used the mean HRI count rate\\
^\ddagger  &  Predicted luminosity corrected for total absorption\\
\end{array}
\]
\end{table*}

To further classify the brighter point sources, we have made use of the spectral
capabilities of the PSPC. For seven sources (X12, X17, X21, X33, X34, X36 
and X40) $\ga 100$~PSPC counts are detected. X33 and X34, 
embedded in the extended nuclear X-ray emission, have already been
discussed in Sect.~\ref{verybright}. 
Photons for the other sources were extracted with extraction
radii of 25$''$, and a local background, determined in a
concentric ring from $r=25''$ to $r=40''$,
is subtracted to reduce contributions from surrounding diffuse emission
features. Contributions from other point sources in our catalog to
this background are avoided as they are screened out with a cut radius of 
25$''$. The very small extraction diameters
and the varying PSF of the individual channels were corrected for using standard
EXSAS procedures.
The counts contained in the raw spectra
are listed in Table~\ref{fits}. Due to the PSPC PSF of $\sim 25''$, we are not 
able to separate X17 from X18 and X40 from X42.
Simple spectral models, a power law (POWL), a thermal
Bremsstrahlung (THBR) and a thin thermal plasma (THPL) model were fitted, and 
for all sources, the formal $\chi^2/\nu$ value of the THBR, THPL and POWL
fits were of the same order of magnitude. In Table~\ref{fits} 
the results of the THBR fits are listed. 
All sources are intrinsically absorbed, and the lowest absorption
($N_{\rm H}\sim 1\times 10^{21}$~cm$^{-2}$) 
is measured for X21, while the fits of all 
other sources indicate $N_{\rm H}\ga 7\times 10^{21}$~cm$^{-2}$. While the
THBR fit predicts a temperature around $1~$keV for X21 the temperatures of the
other sources ($\sim 0.4$~keV) seem to be lower. However, the errors of the
fits (1$\sigma$ errors given in Table~\ref{fits}, 
calculated from error ellipses) are 
high. 
With the help of the more precise HRI count rates and using the
conversion factors relevant to the PSPC suggested models,
we derived two types of source luminosities that are given in Table 7:
an "absorbed luminosity" (i.e. calculating the
flux for temperature and normalisation as suggested by the fit and
using an $N_{\rm H}$ value that is the fit value minus the Galactic
foreground value), $L_{\rm x}$ is
$\sim (1-2)\times10^{38}$~erg~s$^{-1}$ for all the sources;
an "intrinsic luminosity" (i.e. calculating the
flux for temperature and normalisation as suggested by the fit and
using an $N_{\rm H}$ value of zero).
Technically, these intrinsic luminosities are higher
than the absorbed ones by a factor of 2 for
X21 and $\sim 30$ for all other sources. However, especially the big
corrections have to be taken with care as they contain big errors
introduced by the relatively low temperatures and high absorption values
and the associated uncertainties.

\subsection{Emission components of the NGC~253 disk}
\label{integ}
In this section the contribution of the NGC~253 point sources are compared 
to the total X-ray emission of the NGC~253 disk.
One has to be aware that, due to the not 
completely edge-on orientation of the NGC~253 disk, contributions from hot
gas in the lower halo of the galaxy (cf. PEA) will be contained in the 
integral disk count rate. The measured count rates have been corrected
for exposure, deadtime, and vignetting. 
The background was taken from two source
free regions outside the disk and halo of the galaxy, namely a region
east of the galaxy encircled by the sources X68, X69 and X72, and a 
region west of the galaxy encircled by X2, X8 and X20. The PSPC has a
very low detector internal background and enables a sensitive calculation of
count rates over large areas. The results for the HRI, which has a higher
detector internal background, give a higher error and depend more critically
on the background regions chosen. 

\begin{table*}
\caption{Emission components of the NGC~253 disk}
\label{psluminosities}
\begin{flushleft}
\begin{tabular}{llcccc}
\hline
\noalign{\smallskip}
Region & Detector & Count rate & $L_{\rm x}$~$^\star$ 
 & HR1 & HR2 \\
       &          & cts~s$^{-1}$ & $10^{39}$~erg~s$^{-1}$\\
\noalign{\smallskip}
\hline
\noalign{\smallskip}
disk of NGC~253 & PSPC   
&  $0.44$$\pm$$0.01$ & $\phantom{^\dagger}$[$4.6$$\pm$$0.1$]$^\dagger$
& $0.50$$\pm$$0.01$ & $0.18$$\pm$$0.01$ 
\\ 
$\phantom{av}$integral emission
                                 & HRI & $0.15$$\pm$$0.01$ & 
$\phantom{^\dagger}$[$5.0$$\pm$$0.1$]$^\dagger$ 
\\
\noalign{\smallskip}
point sources in the disk 
     & PSPC (28 sources detected) 
  & $0.102$$\pm$$0.002$ & $1.05$$\pm$$0.02$ & $^\ddagger$ & $^\ddagger$ \\
$\phantom{av}$excluding the extended         
                        & PSPC (excluding X50,51,52,57,62) 
  & $0.094$$\pm$$0.002$ & $0.97$$\pm$$0.02$\\
$\phantom{av}$central source X34
                                 & HRI  (25 sources detected) 
  & $0.031$$\pm$$0.001$ & $1.03$$\pm$$0.03$\\
\noalign{\smallskip}
\hline
\end{tabular}
\end{flushleft}
\vskip-.5cm
\[
\begin{array}{lp{0.95\linewidth}}
^\star   & Corrected for Galactic foreground absorption\\
^\dagger & The diffuse emission component detected from the disk
           of NGC~253 cannot be described by a 5~keV thermal Bremsstrahlung
           model (cf. PEA)\\ 
^\ddagger & The hardness ratios of individual point sources are subject of
            Sects.~\ref{hrrr}, and they have been sketched
            in Fig.~8.
\end{array}
\]
\end{table*}

The total count rates within the $D_{25}$ ellipse of NGC~253, together with the
integral point source content 
for the PSPC and HRI observations are given in Table~\ref{psluminosities}. 
The PSPC and HRI count rates for the disk 
are in good agreement
(i.e. the PSPC count rate is higher by a factor of $\sim3$, as expected from 
the energy conversion factors for the different spectral models 
(cf. Table~\ref{ecfs}). 
Excluding the central source X34, which is an extended
source (cf. Sect~\ref{verybright}), the integrated emission of the point
sources makes up roughly one quarter of the total emission.
With both detectors, an integrated point source luminosity of 
$\sim1\times 10^{39}$~erg~s$^{-1}$ is measured. 
The PSPC sources X50, X51, X52, X57 and X62, which are probably due to local
enhancements of the diffuse emission covering the disk of NGC~253, contribute
$1\times 10^{38}$~erg~s$^{-1}$. 

The central source X34 and the surrounding diffuse emission (cf. the lowest
contour level of Fig.~4) contribute (after subtraction of X33 according to 
Sect.~\ref{verybright}) $\sim 0.04$~HRI~cts~s$^{-1}$,
slightly more than the integral point source contribution
($\sim 0.03$~HRI~cts~s$^{-1}$).

\begin{figure}
\label{myhrplott}
\includegraphics[bb=66 156 484 543,width=8.8cm,clip=true]{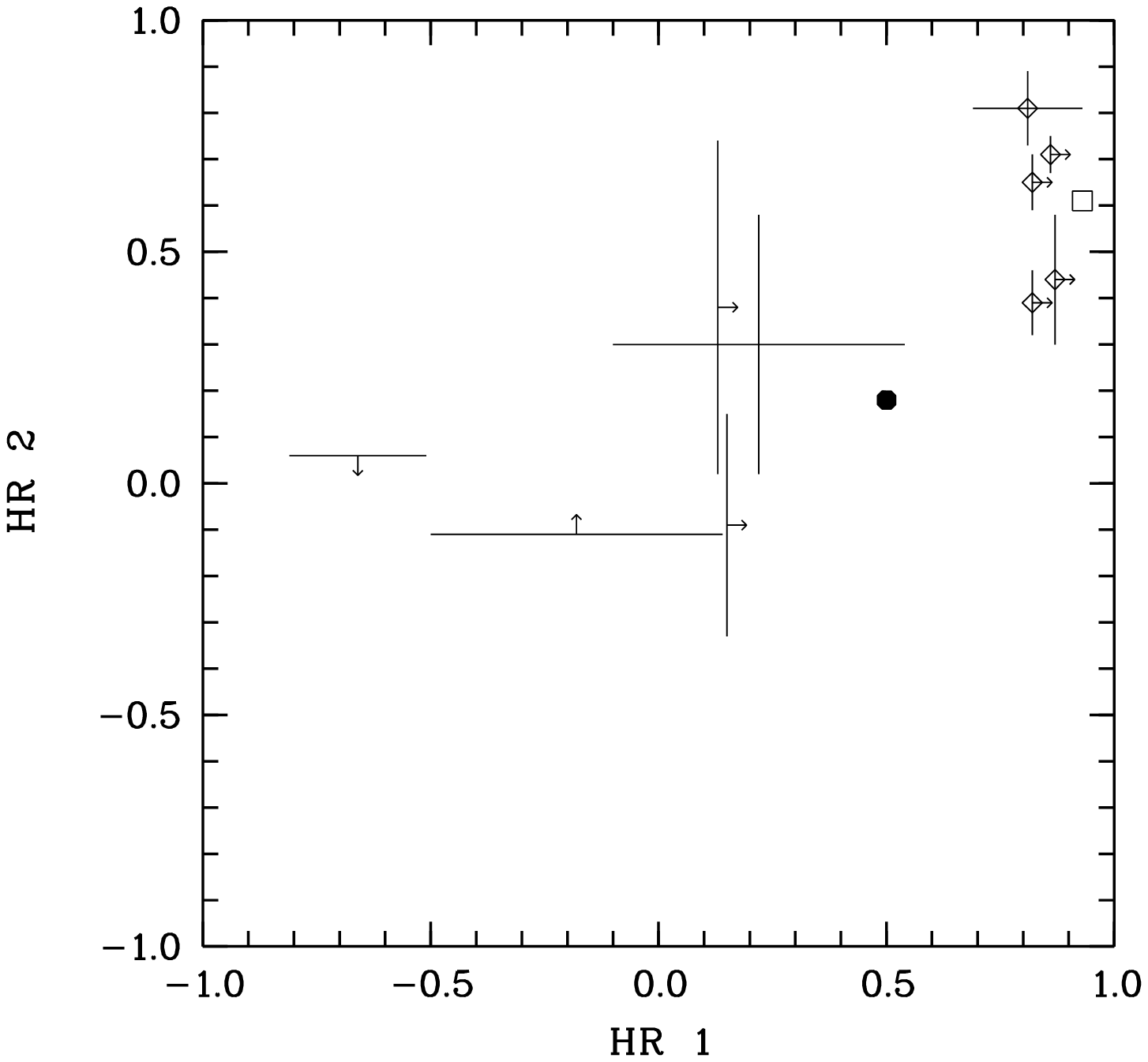}
\caption{ 
Hardness ratio plot of point sources in the NGC~253 disk. The square marks the
position of X33 in the HR diagram. 
The filled hexagon
marks the hardness ratios of the entire NGC~253 disk emission (cf. 
Table~\ref{fits}).
Crosses indicate the measurements and errors for different 
sources in the disk of NGC~253. One group (marked with diamonds at the center
of the crosses) are bright point sources which were  
detected with the PSPC and HRI, the second group (left of the 
hexagon) represents sources only detected with the PSPC and not visible
during the ROSAT HRI observations  
        }
\end{figure}

To visualize the spectral behavior of the different NGC~253 emission 
components, Fig.~8 shows a hardness ratio plot of the total
NGC~253 emission (filled hexagon) and the bright point source X33 (square).
Five further bright PSPC sources (X12, X17, X21, X36, and X40, 
all with $\ga 100$~PSPC~counts), that were also detected with the HRI, have been 
added to the diagram (marked with diamonds). Sect.~\ref{howhard} described,
how HR2 for the sources (all sources were
confused) have been derived. 
Due to the high FWHM of the soft band PSF, the HR1 values are not given
in Table~3. To estimate the HR1 for the purpose
of this section, the soft and hard band counts were calculated within an
extraction diameter of 1.5 times the PSF of the energy bands, and a 
local background from 1.5 to $2.5\times$ FWHM diameter was
subtracted. 
Similar to X33, the sources X12, X17, X21, X36 and X40,
are located in the upper right corner of the diagram, and their high HR1 and
HR2 values suggest harder spectra and higher absorption than for the total
NGC~253 disk emission. The positions of the spurious PSPC sources 
(X50, X51, X52, X57, and X62) in the hardness ratio digram
(left of the hexagon representing the entire NGC~253 disk emission) would
indicate lower absorption and softer spectral behavior, giving further 
support to their identification as spurious detections of diffuse emission 
from the (outer) disk of NGC~253.

Diffuse emission in the disk is investigated further in PEA.
 
\section{Discussion}
Extrapolating from our knowledge of X-ray point-like sources in the Galaxy, 
its neighboring galaxies in the Local Group and 
further nearby galaxies, the sources detected in NGC~253 should be
from the following classes: 
{\bf (1)} 
X-ray binaries. Here, a compact
object (a white dwarf (WD), a neutron star (NS), or a black hole (BH)) 
accretes material from
an accompanying object. Depending on the nature of the
donor star, X-ray binaries are classified as high mass X-ray binaries (massive 
O or B stars as companions) or low mass X-ray binaries (late type stars 
as companions). The maximum X-ray luminosity that can be achieved in such a 
system, assuming steady spherical 
accretion, was calculated by Eddington (1928) as
$L_{\rm x} = 1.3\times 10^{38}$~erg~s$^{-1}\times M/\rm M_\odot$, where $M$
is the mass of the compact object. Only when the compact object 
is a NS or BH is most of this luminosity emitted in X-rays. As NSs have typical
masses of 1 M$_\odot$, the commonly assumed maximum luminosity for a NS system
is $1.3\times 10^{38}$\ergsec. If the luminosity of an X-ray source via the 
Eddington formula indicates a mass
$\ga 3 \rm M_\odot$ -- and the source has to be explained as an X-ray binary -- 
only black holes can act as the accreting objects, as theory does not allow
such masses for WDs and NSs. 
{\bf (2)} 
Supernovae. In the ROSAT band, X-ray emission from supernovae (SNe)
is expected from the interaction between the SN ejecta and the circumstellar
matter, heated by the outgoing wave. 
To date, eight detections of X-ray emitting SNe are reported,
with peak X-ray luminosities in the range $10^{35}$ -- $10^{41}$~erg~s$^{-1}$,
followed by an exponential rate of decline (cf. Schlegel 1995 for a review of
X-ray observations of SNe to the year 1995 and Fabian \& Terlevich 1996;
Lewin et al. 1996; Immler et al. 1998a
and Immler et al. 1998b for detections thereafter).
Based on the estimated SN rate of $1.3 \pm 0.3$ per century per
$10^{10}~L_{\rm B}$($\odot$) for galaxies of type Sc (van den Bergh 1993)
and a total blue luminosity  of NGC 253 of
$1.1 \times 10^{10}~L_{\rm B}(\odot)$ (Tully 1988), a SN rate of $\sim$~1--2
per century is expected within NGC 253, these sources, by now, possibly 
having evolved into X-ray
point sources. It is clear that this simple estimate will only roughly 
describe the SN rate in the disk of the galaxy and not the one connected to the
nuclear starburst activity. Sources close to the nucleus will however 
be too highly absorbed in any event and -- if ever visible at all -- 
unresolvable with the HRI. 
{\bf (3)} Supernova remnants. SN remnants in the Milky-Way 
and our neighboring galaxies attain X-ray luminosities of
up to several 10$^{36}$~erg~s$^{-1}$. A very X-ray bright remnant ($3\times
10^{38}$~erg~s$^{-1}$) is reported for NGC~4449 (e.g. Blair et al. 1983, 
Vogler \& Pietsch 1997). 
{\bf (4)} Super-bubbles. In regions of enhanced
star formation, correlated winds of massive stars ($t\la 10^7$~yr) and 
SN explosions ($t\ga 10^7$~yr) can heat a part of the interstellar medium
to $T\ga 10^6$~K. Super-shells surrounding these bubbles might 
complicate the detection of the hot interior. 
Additional X-ray emission connected to super-bubbles is expected from 
X-ray binaries, 
SNe or SN remnants contained in the star forming regions. 
The brightest known super-bubbles in the LMC 
(the NGC~44 super-bubble and shell 5 in 30 Dor) have luminosities of 
$5\times 10^{37}$~erg~s$^{-1}$, and in M~101, five bright \HII\ regions 
have luminosities from $ 1\times 10^{38}
$~erg~s$^{-1}$ to $3\times 10^{38}$~erg~s$^{-1}$
(Chu \& Kennicutt 1994, Williams \& Chu 1995),
while most of the known super-bubbles have lower luminosities. 

In the Milky-Way, SN remnants and super-bubbles can be spatially resolved with 
the ROSAT instruments. At the distance of NGC~253, one expects these source
classes to be point-like at the resolution of ROSAT, as long as the spatial
extent is $\la 60$~pc. This extent can only be exceeded in the case of
giant super-bubbles (cf., e.g. the HRI detection of a super-bubble in NGC~3079,
Pietsch et al. 1998b).

\subsection{Comparison of the ROSAT point source catalog with {\it Einstein}}
\label{einst}

\begin{table}
\caption{Comparison of {\it Einstein} and ROSAT detected point sources
         in NGC~253}
\label{einstein}
\begin{flushleft} 
\begin{tabular}{llrrr}
\hline
\noalign{\smallskip}
ROSAT & {\it Einst.}&$L_{\rm x}^{\rm Einst. \ \dagger}$
   &$L_{\rm x}^{\rm HRI \ \dagger}$
   & $L_{\rm x}^{\rm PSPC \ \dagger}$\\
\noalign{\smallskip}
\hline
\noalign{\smallskip}

$\phantom{0}$X9  &---& $\le4.4$&$1.2\pm0.3$ & $0.8\pm0.2$\\
X12 &---& $\le7.7$&$^\star 10.3\pm1.6$& $^\star 6.1\pm0.2$ \\
X14 &---& $\le7.7$&$^\star 14.3\pm2.3$&  $\le0.8$\\
X15 &---& $\le2.5$& $^\star 2.9\pm 1.0$ & $3.2\pm0.5$\\
X16 &---& $\le4.3$& $0.9\pm0.3$ & $\le 0.8$ \\
X17 &E2 & $9.7\pm1.9$ & $^\star 14.0 \pm 2.3$ & $3.2\pm0.4$\\
X18 &---& $\le5.5$ & $2.2\pm0.4$ & $5.2\pm0.9$\\
X19 &---& $\le4.5$ & $2.9\pm0.5$ &$ 2.1\pm 0.2$\\
X21 &E5 & $23\pm3.0$  & $11.2\pm0.8$ & $7.3\pm0.6$\\
X23 &---& $\le3.2$ & $2.3\pm0.4$ & $2.5\pm0.4$\\
X25 &---& $\le3.2$ & $2.3\pm0.5$ & $2.4\pm0.3$\\ 
X26 &---& $\le5.9$ & $0.7\pm0.2$ & $\le1.4$\\
X28 &---& $\le2.4$ & $^\star 6.0\pm 1.6$ & $\le1.5$\\
X29 &---& $\le7.7$& $0.8\pm0.3$ & $8.0\pm0.9$\\
X32 &---& $\le2.3$ & $1.1\pm0.3$ &$\le1.2$\\
X33 &E8 & $26\pm3.3$ & $29.5\pm1.4$ &$^\star 30.2\pm1.1$\\
X35 &---& $\le5.8$ &$^\star 4.3\pm1.4$ & $\le1.2$\\
X36 &E1 & $10\pm2.1$  & $8.0\pm0.7$ & $7.2\pm0.6$ \\
X40 &---& $\le7.7$ & $^\star 14.9\pm2.3$ & $9.7\pm0.7$\\
X41 &---& $\le2.7$ & $0.9\pm0.3$      & $\le1.2$ \\
X42 &---& $\le2.2$ & $2.7\pm0.5$      & $2.6\pm0.6$ \\
X44 &---& $\le4.1$ & $0.9\pm0.3$      & $0.9\pm0.2$ \\
X48 &---& $\le2.5$ & $1.9\pm0.4$      & $2.7\pm0.7$ \\
X50 &---& $\le2.3$ & $\le0.7$           & $2.2\pm0.5$ \\
X51 &---& $\le2.7$ & $\le0.2$           & $0.9\pm0.4$ \\
X52 &---& $\le2.8$ & $\le0.8$           & $2.0\pm0.5$ \\
X55 &---& $\le4.9$ & $1.2\pm0.3$      & $1.5\pm0.4$ \\
X57 &---& $\le2.9$ & $\le0.6$           & $1.3\pm0.5$ \\
X62 &---& $\le2.5$ & $\le1.2$           & $1.9\pm0.6$ \\
---&E3 & $12\pm2.5 $& $\le0.9$ & $\le1.7$\\
---&E4 & $13\pm2.6 $ & $\le0.6$ & $\le0.7$\\
---&E6 & $5.6\pm1.4$ & $\le0.8$ & $\le1.4$\\
---&E7 & $5.1\pm1.3$& $\le1.0$ & $\le2.0$\\
\noalign{\smallskip}
\hline
\end{tabular}
\end{flushleft}
\vskip-.5cm
\[
\begin{array}{lp{0.95\linewidth}}
^{\dagger} & Luminosities measured with the {\it Einstein} HRI,
the ROSAT HRI and the ROSAT PSPC. Given
in 10$^{37}$~erg~s$^{-1}$ (0.1--2.4~keV). Assumed spectral model:
5 keV thermal bremstrahlung,
corrected for Galactic foreground absorption. For non-detected sources we
give $2\sigma$ upper limits.\\
^\star    &   Maximum luminosity (cf. Sect.~3.3 and Table~6)
\end{array}
\]
\end{table}

To investigate the long term time variability of the ROSAT detected NGC~253 
point sources,
the ROSAT results were compared with the {\it Einstein} HRI data, collected in
July 1979. The {\it Einstein} source list contains eight 
point-like sources (sources E1$\,$--$\,$E8, 
Fabbiano \& Trinchieri 1984, cf. our Fig.~2 and the figure caption 
for the position of the sources). To calculate upper limits (2$\sigma$) 
to the {\it Einstein} luminosity at the positions of 
({\it Einstein}-undetected) ROSAT sources, {\it Einstein} data have been 
retrieved from the High Energy Astrophysics
Science Archive Research Center (HEASARC), operated by the 
Goddard Space Flight Center (GSFC). Table~\ref{einstein} 
compares luminosities and upper limits for the {\it Einstein} and
ROSAT PSPC and HRI detected sources. 

27 sources are detected within the $D_{25}$ ellipse NGC~253
by ROSAT, 9 of which show time variability within the ROSAT observations
(cf. the results in Sect.~\ref{disksource}). For four of the time
variable sources (X12, X14, X28 and X40) 
the {\it Einstein} upper limits lie below the ROSAT measurements, and
this further strengthens the idea that these sources are time variable.
Two additional time-variable ROSAT sources (X17, X21) were also detected 
with {\it Einstein}, and for these sources, the 
different {\it Einstein} and ROSAT luminosities support the picture of time 
variability. Two ROSAT sources with no known time variability 
were detected with {\it Einstein}, 
namely the bright point source close to the nucleus
(X33) and X36. ROSAT and {\it Einstein} luminosities for these sources 
agree within the errors, and no time variability can be inferred. 
Due to the reduced sensitivity of the {\it Einstein}
observation however 
(detection limit $\sim 2.5\times 10^{37}$~erg~s$^{-1}$, compared
to $\sim 7\times 10^{36}$~erg~s$^{-1}$ for ROSAT), 
we cannot make any long term time variability arguments for the 
remaining fainter ROSAT sources. Conversely, 
four sources within or close to the $D_{25}$ ellipse of NGC~253 were seen
with {\it Einstein}, but not with ROSAT: E3, E4, E6 and E7. These
non-detections with ROSAT (the ROSAT upper limits being significantly below
the {\it Einstein} luminosities) argue for the detection of 
transients in the case of these sources. 

\subsection{The nature of the NGC~253 point sources}
\label{sothisisthenature}
The nature of the sources associated with the NGC~253 disk is discussed in this
section. As will be shown in Sect.~\ref{forback}, only a negligible number
(of order 1) of foreground or background X-ray sources is expected within
the $D_{25}$ ellipse of NGC~253.

Combining the ROSAT and {\it Einstein} observations, a total of 31
point sources in NGC~253 are seen. Two  {\it Einstein} sources, 
E3 and E4, located close to the D$_{25}$ ellipse of NGC~253, 
are included in this
number. This seems justified though, as {\sc Hi} observations of NGC~253 
(Puche et al. 1991) and deep optical observations (Beck et al. 1982)
indicate an extent of NGC~253 far beyond the D$_{25}$ ellipse. Also, their 
transient nature (see Sect.~\ref{einst}) strengthens the idea that
E3 and E4 are members of NGC~253, as had already been proposed by Fabbiano \& 
Trinchieri (1984).

Time variability is detected in 13 of the 31 NGC~253 sources, and nearly one
half of these (X12, X14, E3, E4, E6 and E7) show transient behavior. 
X12 and X14 reach their highest luminosities during the ROSAT observation
blocks 4 and 6, respectively. Compared to the lowest upper limits 
in the case of non-detections (block 7 for X12 and block 4 for X14), the
peak luminosity is higher by 4.0$\sigma$ and 5.6$\sigma$ ($\sigma$ 
represents the measurement error of the peak luminosity) for X12 and X14,
respectively. While X14 is not detected with the PSPC, X12 is
detected during block 3. The luminosity is smaller than the one
measured with the HRI however, due to the smaller error of the PSPC 
measurement, the PSPC 
peak luminosity is 19.8$\sigma$ above the upper limit for block 7.
In the case of the {\it Einstein} detected transients E3, E4, E6, and E7, the
significances of the luminosity above the ROSAT upper limit are 4.4,
4.3, 3.4, and 3.2$\sigma$, respectively. 

The interpretation of the bright time-variable sources is 
relatively straightforward.
Taking into account the fact that -- with the exception of SN 1940E -- no SNe 
have been reported in NGC~253, the time-variable sources have to 
be classified as X-ray binaries.
With the exception of X33, the derived maximum luminosity of the
variable sources 
($1.5\times 10^{38}$~erg~s$^{-1}$) and the variability
could be well explained if one assumes X-ray binary systems containing accreting
objects of mass $M\sim 1\,\rm M_\odot$ ($e.g.$ neutron stars) radiating
close to the Eddington limit. However, if additional absorption, intrinsic 
to the sources or from the \HI\ disk 
of NGC~253 is included, higher intrinsic luminosities of the sources 
are implied. An additional $N_{\rm H}$
of $1\times10^{21}$~cm$^{-2}$ or $5\times10^{21}$~cm$^{-2}$ 
(cf. the \HI\ map presented in Puche et al. 1991), for example, would 
imply source luminosities higher by factors of 1.6 and 4, 
respectively. Under these assumptions, the maximum X-ray luminosities
of all the mentioned binary candidates would be close to, or even  exceed 
the Eddington limit for a neutron star binary. 
If one assumes that X33 is an X-ray binary radiating at the Eddington 
limit, a mass for the compact object of $\ga 3\,{\rm M}_\odot$ can be 
calculated, that clearly puts the object in the mass range expected for a 
black hole.

It is more difficult to determine the nature of the remaining, 
less bright X-ray sources in NGC~253. Firstly,
the non-detection of time variability with ROSAT might be due to the low 
photon statistics or the time windows of the observation blocks, and, in 
general, one cannot rule out the idea 
that the luminosity of these sources is variable. 
Secondly, none of the X-ray positions coincide with point sources visible 
in radio maps that could be due to SNe, SNRs or \HII\ regions.
This may not be too astonishing, bearing in mind that the brightest supernova 
remnant in the LMC (N158A) has 
$L_{\rm x} = 5\times 10^{36}$~erg~s$^{-1}$ (Chu \& Kennicutt 1994), 
well below the detection limit of the ROSAT observations for NGC~253. 
Therefore, one might again argue in favor of an X-ray binary identification.
Another explanation one could put forward is that unresolved emission from 
SNe and SN remnant and X-ray binaries, embedded in the hot interstellar
medium of star forming regions, would suppress time variability. 
Such a scenario might also explain 
the bright source X36 (no time variability established)
with a luminosity 
($L_{\rm x} \sim 1\times 10^{38}$~erg~s$^{-1}$)  comparable to that
of X-ray bright super-bubbles reported in M~101 (Williams \& Chu 1995).

In contrast to face-on galaxies like M~101, the edge-on
orientation of NGC~253 complicates the detection of \HII\ regions in 
H$\alpha$ observations, and only for X42 ($L_{\rm x} = 3\times 10^{37}
$~erg~s$^{-1}$) could a positional coincidence be established
(reported by Waller et al. 1988).

\subsection{Non-detection of SN 1940E?}
The type I supernova SN 1940E (cf., e.g. Barbon et al. 1989) 
is the only historical SN reported in NGC~253. The SN is located 71$''$ west
and 17$''$ south of the nucleus, and its position is marked with a
cross ($\times$) in Fig.~2. Our X-ray catalog contains no source close to
the position of the SN. We searched for faint emission at the position 
of SN 1940E, by extracting HRI counts from a ring of radius 10$''$.
The background was taken from an annulus of 10$''$ to 15$''$ radius.
This procedure resulted in a residual count rate of 
(2.3$\pm$1.3)$\,\times\,$10$^{-4}$~cts~s$^{-1}$ 
(converting to $L_{\rm x}= (7.4$$\pm$$4.2)\times 10^{36}$~erg~s$^{-1}$).
It is not clear, however, whether this 1.8$\sigma$ excess is really emission 
from the SN or is just produced by the patchiness of 
the diffuse X-ray emission in the inner spiral arms of NGC~253.

To date, only one possible X-ray detection of a SN of type I soon after
the outburst has been reported (SN 1994I in M~51,
$L_{\rm x} = 1.6 \times 10^{38}$~erg~s$^{-1}$, Immler et al.
1998b). Comparing with this paper, and 
taking the luminosity upper limit above and 
$L_{\rm x} = \int \Gamma(T) (4\rho)^2 dV$ with a cooling function
of $\Gamma(5~{\rm keV}) = 3 \times 10^{-23}$ erg cm$^3$ s$^{-1}$
(Raymond et al. 1976), one can estimate a mean density of
$\rho < 30~{\rm cm}^{-3}~v_{10\,000}^{2/3}$ 
(for a shell expansion velocity in units of 10\,000 km s$^{-1}$)
and a total mass of X-ray luminous gas of $M < 0.5 M_{\sun}$ inside
a sphere of radius $\sim 10^{19}$~cm. While this density limit is typical for 
the interstellar medium in the disk of galaxies, it is four orders
of magnitude lower than that expected for the gas deposited by type I SNe due 
to non-conservative mass transfer to a companion or due to stellar wind prior
to the outburst (cf. Immler et al. 1998b). There are
several ways to explain this discrepancy: SN 1940E may be embedded deep in 
the NGC~253 disk, and the count to luminosity conversion may underestimate this
effect, the assumptions for the cooling may be wrong, leading to a too high 
luminosity after 55 years, or the emission from
SN 1994I may not be typical for type I SN.  

\subsection{Possible contributions from foreground or background sources}
\label{forback}
We have attributed all X-ray sources found within the optical extent of 
NGC~253 to the disk of NGC~253 
(with the exception of X58, a background object, cf. appendix A). 
We demonstrate that this assumption is justified, in the following, by proving 
that only a negligible number of
foreground and background sources is expected in this area.

Firstly, one can estimate the number of X-ray sources due to foreground objects
by extrapolating the local density of X-ray detected stars in the field. To do 
so, only HRI sources were taken into account. In this way we avoided PSPC 
sources that might partly represent diffuse emission in the halo of NGC~253.  
From the 34 HRI sources outside the $D_{25}$ ellipse
of NGC~253, only the sources X31 and X61 are identified with foreground
objects (stars, see appendix A). 
Keeping this in mind and comparing the area covered by the disk of 
NGC~253 ($120$~arcmin$^{2}$) to that of the remaining HRI field of view 
($\sim 950$~arcmin$^{2}$), one expects 0.25 foreground source 
within the disk of NGC~253, making the detection of such a source 
rather unlikely. 

Secondly, one can estimate the number of background sources shining through 
the disk of NGC~253 in a similar way. 
We can assume conservatively that all objects outside the NGC~253 disk with the 
exception of X31 and X61 are background objects. 
In addition, we have to keep in mind that the source flux for
objects behind the NGC~253 disk will be reduced due to the additional 
absorption of X-rays that these objects will suffer from the 
interstellar medium within the disk of NGC~253.
While the detection limit for field sources in the HRI field was  
$\sim 1\times 10^{-14}$~erg~s$^{-1}$~cm$^{-2}$, we only expect to detect
sources from behind the NGC~253 disk if they have intrinsic fluxes of at least 
$1.6\times 10^{-14}$~erg~s$^{-1}$~cm$^{-2}$ (correcting for a
typical column density $\ga 10^{21}$~cm$^{-2}$ (Puche et al. 1991)).
Only 18 background sources were detected above this limit outside the NGC~253 
disk. From this and the
ratio of the NGC~253 disk/outside areas, one can predict
2.3 background sources within an area covered by the NGC~253 disk. 
With the help of optical spectroscopy (see appendix), we have already 
identified one background 
QSO, X58, on the border of the $D_{25}$ ellipse of NGC~253.

Another approach to determine the contamination of NGC~253 disk sources with 
background objects uses the deep field luminosity functions derived
in the Lockman hole (Hasinger et al. 1991, 
Hasinger et al. 1993). If we again correct for an average 
$N_{\rm H}$ of $10^{21}$~cm$^{-2}$ within the the disk of NGC~253, 
1.6 background sources are predicted that should be detectable shining through 
the NGC~253 disk. This number is -- within the statistics -- consistent with
the one derived from the field objects. 

\subsection{Comparison to results previously
            published on ROSAT and ASCA observations}
A sample of ROSAT PSPC observed spiral galaxies -- including the 22.9~ks PSPC 
observation of NGC~253 -- has been homogeneously 
analyzed by Read et al. (1997) to search for point source and diffuse emission 
components. In the NGC~253 field, they detect 15 point sources 
(R1$\,$--$\,$R15), seven of which are located within the disk of NGC~253
(i.e. R4 corresponding to the source cataloged above as X12, 
R6/X15, R7/X17+nearby source
X18, R8/X21, R11/X34+nearby source X33, R12/X36 and R13/X40+nearby source 
X41). Due to their small separations,
the sources X17/18, X33/34 and X40/42 are not resolved as individual sources
within the PSPC data. 
The count rates given by Read et al. (1997) for the sources R4$\,$--$\,$R13 
are slightly higher ($\sim 10\%-20\%$) than our rates. 
This can be understood if one keeps in mind that we used a multi source
fit technique to calculate the count rates, and excluded contributions from
nearby sources, which otherwise might increase the count rates. 
In addition, due to a more sensitive source search, our catalog contains 
more sources.   

Ptak et al. (1997) report on ASCA observations of NGC~253. Because of the large
PSF of ASCA (FWHM $\sim 3'$), point sources can not be spatially separated
from diffuse emission components. However, because of the good spectral 
resolution of the detectors, they were able to fit multi component 
spectral models to the integral emission of NGC~253. When comparing with
ASCA, one has to keep in mind that due to their
geometry the ASCA detectors do not cover 
the entire NGC~253 disk, and therefore ASCA count rates have been only 
been extracted 
from a circle of 6$'$ radius around the center of NGC~253. Also, the ASCA 
energy coverage (0.5--10~keV) differs from that of ROSAT. 
In the overlapping 0.5--2.0~keV band,
Ptak et al. report a flux of $2.8\times 10^{-12}$~erg~s$^{-1}$ for NGC~253. 
Extracting from the same area one obtains a count rate of 0.33~cts~s$^{-1}$
in the corresponding ROSAT hard band. 
To be independent of errors introduced by different spectral models, 
the same spectral parameters as used for the ASCA data are used
to convert the ROSAT count rate
to a luminosity. Ptak et al. (1997) fit the entire emission of NGC~253 in the
0.5--10~keV ASCA band as a combination of a thin thermal plasma and a
power law component. They obtain $N_{\rm H}= 1\times 10^{21}$~cm$^{-2}$ 
and $T= 0.8~$keV for the thin thermal plasma component, and 
$N_{\rm H}= 1.4\times 10^{22}$~cm$^{-2}$ and a power law index of 2.0
for the higher absorbed power law component with unabsorbed luminosities 
(0.5--2.0~keV) of 
$2.3\times 10^{39}$~erg~s$^{-1}$ for the thin thermal plasma component and
$3.3\times 10^{39}$~erg~s$^{-1}$ for the power law component. From this,
the calculated luminosities corrected for Galactic absorption are 
$2.0\times 10^{39}$~erg~s$^{-1}$ and 
$5\times 10^{38}$~erg~s$^{-1}$ for the thermal plasma and power
law component, respectively. 
For the same spectral model, the ROSAT 
H band count rate translates to
$L_{\rm x} \sim 2.7\times 10^{39}$~erg~s$^{-1}$, a luminosity 
in very good agreement with the one measured by ASCA. If one still wants to 
explain the slightly  higher
ROSAT luminosity (by $0.2\times 10^{39}$~erg~s$^{-1}$), 
one can argue in terms of the existence of time-variable sources
which might have been picked up with ROSAT  during the six observation blocks 
distributed over several years but not with ASCA (only one observation block). 
In this way, the transient
X12 contributes an average luminosity of $\sim 5 \times 10^{37}$ to the
luminosity determined in the ROSAT PSPC hard band.

\subsection{Comparison to other wavelengths}
We compared our ROSAT point source catalog to images taken at other 
wavelengths to identify possible supernova remnants, H{\sc i} holes and 
H{\sc ii} regions within the NGC~253 disk. 

Radio maps from 0.3~GHz -- 4.7~GHz 
(Carilli et al. 1992, Beck et al. 1994) show a bright 
nuclear source surrounded by diffuse radio emission. The diffuse emission
covers the entire NGC~253 disk and protrudes from the disk into the halo of
the galaxy. On top of the diffuse radio emission covering the bulge, disk and
halo of NGC~253, several enhancements in the radio emission are visible. 
Within the disk of NGC~253, no X-ray sources are found at the position of this 
enhanced radio emission, with the exception of the nuclear area. VLA 
observations of NGC~253 (e.g. Ulvestad \& Antonucci 1997) resolve the central 
$10''$ nuclear region of NGC~253 at a spatial scale of $\sim 1$~pc. These radio
detected SN remnants and \HII\ regions can be attributed to the
starburst nucleus of NGC~253. However, the resolution of the
ROSAT data ($\sim 60~$pc) is insufficient to identify (a part) of the 
extended central X-ray emission with individual radio point sources. 
On the other hand, the enhanced radio emission coincides nicely with 
the bright near-infrared emission of the starburst nucleus (Sams et al. 1994);
also the young, luminous, compact stellar clusters
detected with the WFPC2 camera on the Hubble Space Telescope (Watson et al.
1996) are located in the same area. The maximum of the
central extended X-ray source (X34, cf. Sect.~\ref{dieguteanmerkung}) is found
with an offset of $\sim 4$\arcsec\ to the southeast from the
cluster of these bright nuclear radio sources. The fact that this offset is
slightly exceeding the systematic position errors, may indicate that the
nuclear extended X-ray emission is caused by the hottest
part of the gas outflowing from the nuclear region and not
by a collection of individual point sources
(see further discussions in PEA).
                   
The X-ray point source catalog can be compared with fainter 
compact radio sources in the disk by making use of the reanalysed 
6 and 20 cm VLA data (Ulvestad \& Antonucci, in preparation).
From a diameter of 12\arcmin\  they report 
27 compact sources in the NGC~253 disk (outside the nuclear starburst)
and 5 sources close-by. 
While several sources within this new catalog coincide with source
positions that we have already derived from the radio images of 
Carilli et al., no coincidences with the X-ray catalog are seen. This is
slightly surprising as one might have expected that some of these SNRs or \HII\
regions detected in radio would also be bright enough to be detected in X-rays.
An explanation for this behavior may be that, due to the edge on view
on NGC~253, the X-ray emission of these radio sources (mostly soft for this
class of emitters) is heavily absorbed and therefore not detected. 
 
A comparison of the X-ray source catalog to 
H$\alpha$ images (Waller et al. 1988) suggests a coincidence 
of X42 and a bright \HII\ region. Therefore X42 may represent emission of hot
gas connected with the \HII\ region. This would be consistent with the
non-detection of X-ray variability and also the X-ray luminosity of the source.
The hardness ratio HR2 of 0.2, on the other hand, suggests a hard
spectrum, that would not be expected for this class of source.   

\subsection{Comparison to X-ray point sources in other spiral galaxies}
\label{lognlogskap}
After discussing of the nature of the individual point sources in NGC~253,
we may compare the point source population of NGC~253 
with results from other spiral galaxies.

\begin{table*}
\caption{Comparison of the NGC~253 point source content 
         to other spiral galaxies}
\label{noka}
\begin{flushleft} 
\begin{tabular}{llccccccccl}
\hline
\noalign{\smallskip}
Galaxy & Type$^\star$ & Incl.$^\star$ & 
$d$ & $N_{\rm H}$$^\dagger$ & Det. & Det. lim.  
 &point sources 
& $L_{\rm x}~^\sharp$ & $L_{\rm x}/L_{\rm B}$$^\heartsuit$ & Ref.\\
\noalign{\smallskip}
& & $^\circ$ & Mpc & $10^{20}$ & & $10^{36}$   
 & & $10^{39}$ & $10^{-5}$ \\
& & & &~cm$^{-2}$ & & erg~s$^{-1}$&  & erg~s$^{-1}$\\
\noalign{\smallskip}
\hline\noalign{\smallskip}
N~253 & Sc & 86 & 2.58 & 1.3 & H$^\ddagger$& 7  & 25$\,$+$\,$nucl. & 1.0
        & 2.2 & this work\\
%NGC~4258 & Sbc & 71 & $\phantom0$6.4$\phantom0$ 
%  & 1.2 & ROS H & 13 + nucleus & 2.8 & 3.1 & Vogler 1997\\
M~31 & Sb & 78 & 0.69 & 3.0 & P$^\ddagger$ & 0.4 &  
340$\,$+$\,$bulge & 1.8/3.2$^\diamond$ 
& 1.7/2.9$^\diamond$ & Supper et al. 1997\\
M~33 & Scd & 56 & 0.80 & 6.0 & H$^\ddagger$ & 6 & 26$\,$+$\,$nucl. &0.8 & 4.8 
& Schulman \&\\
     &      &   &     &      &   &   &              & & & Bregman 1995 \\ 
%M~51 & Sbc & 64 & $\phantom0$7.7$\phantom0$ & 1.3 & ROS H 
%& 17 + nucleus & 5.1 & 4.3 & Immler 1996 \\
%M~81 & Sab & 60 &   $\phantom0$3.5$\phantom0$ & 3.8 & {\it Einst} H 
%     & 8 + nucleus & 2.5 & 3.2 & Fabbiano 1988\\
%M~100 & Sbc & 37 & 17.1$\phantom0$ & 2.3& ROS H & 9 + nucleus
%      & 7.3 & 4.2 & Immler et al. 1998\\
%M~101 & Scd & $\sim0$ & $\phantom0$7.5$\phantom0$ & 
% 1.1 & ROS H & 21 + nucleus & 6.0 & 3.4 & Immler 1996\\
\noalign{\smallskip}
\hline
\end{tabular}
\end{flushleft}
\vskip-.5cm
\[
\begin{array}{lp{0.95\linewidth}}
^\star    &   Tully (1988)\\
^{\dagger} &  Galactic foreground absorption (Dickey \& Lockman 1990)\\
^{\ddagger} & H = ROSAT HRI, P = ROSAT PSPC\\
%              {\it Einst} H = {\it Einstein} HRI\\
^\sharp    &  0.1--2.4~keV band luminosity of point sources (excluding the
              bulge of M~31 and the sources at the position of the nucleus
              for all other galaxies), corrected for Galactic foreground\\
^\heartsuit & Blue luminosities according to Tully (1988), corrected for the
              given distances and adjusted for reddening\\
^\diamond & No active galactic nucleus or plume like emission in the center
             region has been established for M~31 
             from the {\it Einstein} HRI observations. Assuming that 90\% 
             of the bulge luminosity ($1.6\times 10^{39}$~erg~s$^{-1}$) are
             caused by point sources, one obtains 
             $L_{\rm x}^{\rm ps} = 3.2 \times 10^{39}$~erg~s$^{-1}$
             and $L_{\rm x}^{\rm ps}/L_{\rm B}\sim 2.9
             \times 10^{-5}$\\
\end{array}
\]
\end{table*}

\begin{figure}
\label{lognlogs}
\includegraphics[bb=73 187 612 621,width=8.8cm,clip=true]{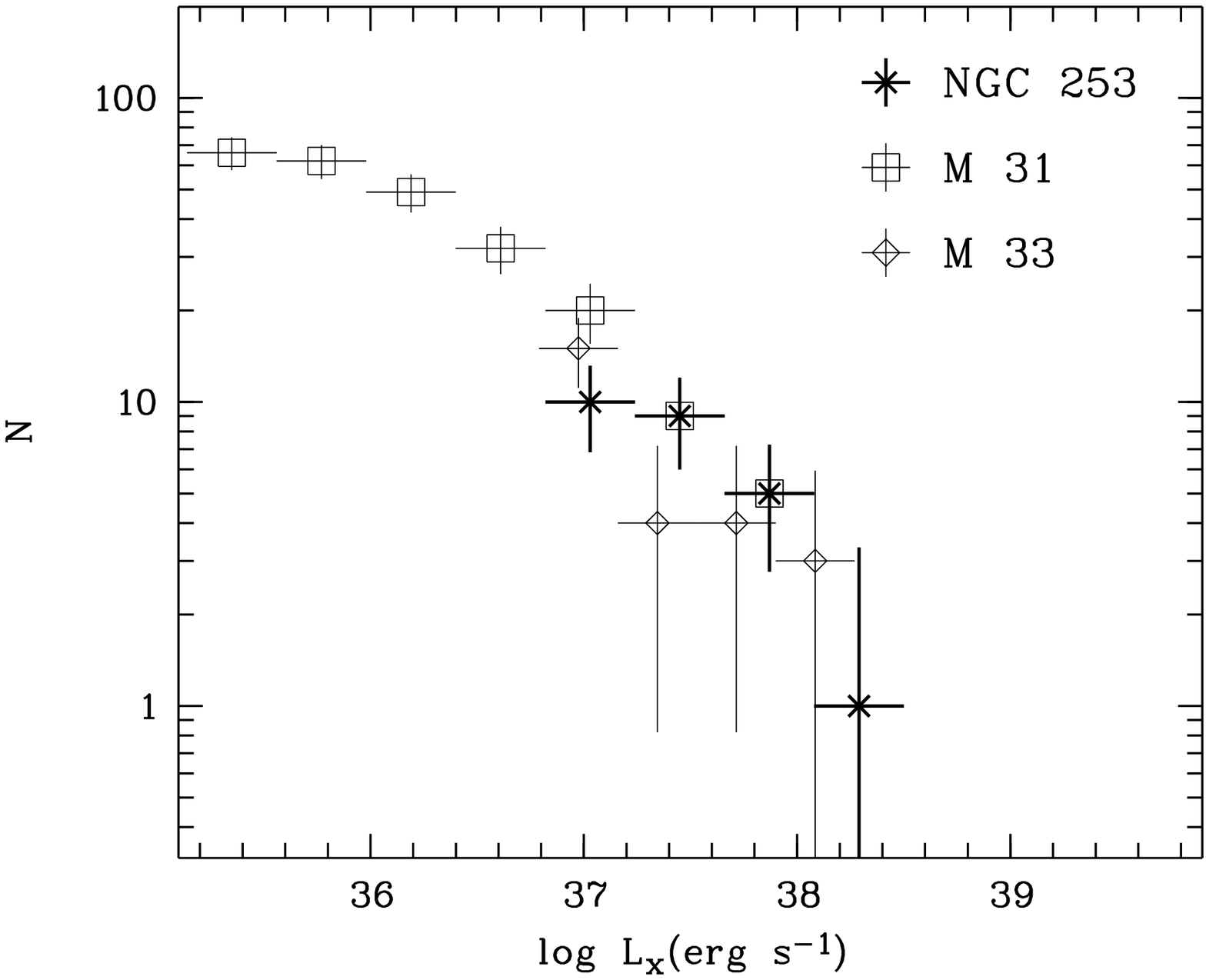}
\caption{
Luminosity distributions of detected point sources (excluding nuclear sources)
for the spiral galaxies NGC~253 (ROSAT HRI), M~31 (ROSAT PSPC, Supper
et al. 1997) and M~33 (ROSAT HRI, Bregman \& Schulman 1995).
The error bars (1$\sigma$) for the individual points have been calculated from
either a Gaussian error distribution or Poisson statistics when appropriate
(cf. Gehrels 1986)
        }
\end{figure}

The local group spiral galaxies M~31 (Supper et al. 1997) 
and M~33 (Schulman \& Bregman 1995) have been 
investigated to
luminosities below that reached for NGC~253 (cf. Table~\ref{noka}). 
We compared the integrated point source content (cf. Table~\ref{noka}) 
as well as the luminosity distributions of the individual
point sources of these galaxies (cf. Fig.~9) to NGC~253. 
The whole optical extent of M~31 has been observed with a
ROSAT PSPC raster scan. In the central area of M~31
($r=1$~kpc corresponding to 5$'$) the PSPC source confusion is very high. 
While this central region was therefore excluded from the luminosity
distribution, the integrated point source content of the region was estimated, 
assuming that 90$\%$ of the bulge luminosity is made up of point sources.
This assumption seems to be justified by {\it Einstein} HRI and ROSAT HRI data
(Trinchieri et al. 1988; Primini et al. 1993), 
which do not show diffuse or plume like emission as in the case of the
NGC~253 nuclear region. 
The ROSAT HRI observation of M~33 was centered on the nucleus of the galaxy 
(Schulman \& 
Bregman 1995). Due to the large optical extent of M~33 ($D_{25}$ ellipse
of $56\farcm5\times 35\farcm0$, Tully 1988), the HRI observation (field of
view $\sim 37'$) did not cover the entire outer disk of the galaxy. 
At the center of M~33 is a bright X-ray source ($L_{\rm x} = 
1.3\times 10^{39}$~erg~s$^{-1}$), its true nature not yet being clear.
The source might be, e.g, a mildly active nucleus. 
To reduce the contributions of active or starburst nuclei to our
point source comparison the central sources of M~33 and NGC~253 (the
extended source X34) were excluded from the total point source luminosity and
the luminosity distribution diagram. However, we are aware that the nuclear
area of M~33 could contain a sample of bright X-ray point sources instead
of a mildly active nucleus, and that the total point source luminosity as
well as the luminosity distribution of M~33 as assumed by us 
would have to be corrected. 

In the case of NGC~253, M~31 and M~33 the ratio between X-ray and optical 
luminosity, which is independent of distance, has been
compared  (cf. Table~\ref{noka}), and it differs slightly, being lowest for the 
nearly edge-on galaxy NGC~253 ($2.2\times 10^{-5}$) and highest for the 
galaxy with the lowest inclination, M33 ($4.8\times 10^{-5}$). One can 
to first order correct for the inclination effect by correcting for
an average additional absorption of 
the NGC~253 sources of $2\times 10^{21}$~cm$^{-2}$ (cf. the {\sc Hi}
map presented in Puche et al. (1991)), resulting in 
$L_{\rm x}^{\rm ps}/L_{\rm B} = 3.4\times 10^{-5}$, lying between the values
found for M~31 and M~33. The ratios only differ by a factor of less than 2 
in spite of the differences in morphological type and star forming 
activity of the galaxies. These results fit nicely to the close correlations 
for X-ray and optical luminosities of spiral galaxies reported by 
Fabbiano et al. (1992) from an analysis of the {\it Einstein}
 sample of galaxies.

The number of point sources detected in different luminosity ranges for 
NGC~253, M~31 and M~33 are presented in Fig.~9. 
The nuclei of NGC~253 and M~31, as well as the inner
bulge region of M~33 have been excluded for the reasons mentioned above. 
As the X-ray sources in M~31 are complete for 
$L_{\rm x}\ga 10^{36}$~erg~s$^{-1}$ (cf. Supper et al. 1997), 
the M~31 curve above this limit defines the luminosity function of point 
sources. Below this threshold it is unclear whether the slower
increase in the number of sources per energy bin is due to a change in 
the point source population or due to incompleteness. 
Within the errors, the luminosity
distributions of the NGC~253 and M~33 point sources 
match within the overlapping region, to that of
M~31. This suggests that source populations for the disks  of these galaxies
(excluding the nuclear regions) are similar. 
This result is somewhat surprising, as one might have expected an enhanced 
number of point sources connected with the star forming activity of NGC~253 that
has not only been reported from the nuclear starburst region but also 
from the boiling galactic disk (Sofue et al. 1994). 
The present measurements do however, not totally rule out this
possibility, as there could be an additional contribution of lower luminosity
point sources that are not detected due to the ROSAT sensitivity cut-off. 

In a couple of more distant spiral galaxies very bright 
($L_{\rm}\ga 3\times 10^{39}$~erg~s$^{-1}$), non-nuclear 
point-like sources were detected by ROSAT, namely in 
NGC~891 (Bregman \& Pildis 1995), 
NGC~4559 (Vogler et al. 1997), NGC~4565 (Vogler et al. 1996),
NGC~6946 (Schlegel et al. 1994) and M~100 (Immler et al. 1998a). In the case of
NGC~891 and NGC~6946, the X-ray emission is due to SNe. In the case of the
other galaxies, the nature of the bright sources is still unclear. The
sources are point-like at the resolution of the ROSAT HRI (NGC~4559, M~100)
and PSPC (NGC~4565, no HRI observation available). 
The non-detection of recent SN outbursts in these galaxies favors the
explanation of X-ray binary systems or SN remnants 
expanding into high density ($n\ga 10^{3.5}$~cm$^{-3}$) media. 
In the case of X-ray binaries, these objects would most likely have to
be black hole binary candidates, 
the masses of the central objects exceeding $10\, \rm M_\odot$. 
However, the sources could also be due to a
superposition of X-ray binaries and SN remnants, as expected in super-bubbles.
Assuming these bright sources are really point-like, one would
expect, contrary to the observations, that such sources should also exist
in NGC~253, M~31, and M~33. 
Their absence can either be explained in terms of time variability, 
where no such source was active during the ROSAT
observations, or by a differing source population. 

\subsection{Point sources outside the disk of NGC~253 possibly correlated
            to the galaxy}
While point sources located inside the $D_{25}$ ellipse of NGC~253 can 
fully be attributed to the galaxy (with the exception of one or two
background sources, cf. Sect.~\ref{forback}), it is more difficult
to decide whether point sources located (projected) in the halo 
of NGC~253 are associated with the galaxy. Such a correlation could be expected
for globular clusters, which might contain low mass X-ray binaries,
or for sources due to emission from hot gas in the halo.

The X-ray point source list is compared to 
globular cluster candidate lists based on optical observations 
(Liller \& Alcaino 1983, 63 candidates; Blecha 1986, 32 candidates). 
No positional coincidence can be established, and this seems to contradict 
expectations deduced from ROSAT observations of M~31. 
Of the ROSAT sources in M~31, 43 have luminosities above the
detection limit of NGC~253, and more than half (26) of 
these bright sources have been identified as globular clusters.
Assuming a similar ratio for NGC~253, 16 
sources would be expected to be globular cluster sources. 
However, a direct comparison between the two galaxies is rather difficult.
$90\%$ of the globular clusters detected in X-rays in M~31 are located
within an `inner disk' described by an ellipse half the size of M~31. 
A list of optical candidates for M~31 globular clusters in this
region could be obtained because of the lower inclination (78$^\circ$ versus
86$^\circ$ for NGC~253) and the smaller distance (0.69~Mpc versus 2.58~Mpc).
Such a list is not available for the inner disk of NGC~253. 
In addition, X-rays from only one globular cluster candidate outside the 
area covered by the M~31 disk has been detected at a distance of 
$\sim 6$~kpc from the galactic plane. 
For NGC~253, however, most of the globular cluster candidates
contained in the lists of Liller \& Alcaino (1983) and Blecha (1986) are 
located at large distances from the plane of the galaxy ($d\ga 5$~kpc
for nearly all candidates in the list of Liller \& Alcaino 1983).

An example of a source in the NGC~253 halo, caused by diffuse emission, and 
having a point-like appearance at the resolution of the PSPC (and possibly 
also at the resolution of the HRI) could be, e.g., a region 
of (older) dense interstellar medium in the halo that is shock-heated by a 
superwind (cf. e.g. Suchkov et al. 1994). Seven sources in the halo of NGC~253 
are surrounded by diffuse X-ray emission attributed to hot gas,
namely the sources X10, X13, X22, X24, X27, X30
and X45. Faint optical objects close
to the HRI and PSPC detected sources X13 and X22 (separation $ < 2''$)
 suggest background AGNs as
sources of the X-rays. The faint object at the position of X22 could be
spectroscopically identified as QSO (see appendix A).
In the case of X27, only detected with the PSPC, a faint object is found 
at a distance 12$''$ (less than the position error) on the ROE finding charts. This might suggest an
identification with a background AGN. 
The non-detection of this source with the 
HRI might indicate time variability. The optical object, however, has not
yet been spectroscopically identified. Taking into account the larger 
separation between the optical and X-ray position of X27,
 compared to X13 and X22, 
and that, in addition, only X27 was not detected with the HRI, one 
alternatively might argue that X27 is a diffuse emission feature. 

No optical candidates
were found in the cases of X10, X24 (both only detected with the PSPC), X30 
(only detected with the HRI) and X45 
(detected with the HRI and PSPC). In the case of X30, 
time variability was established
with the help of the HRI observation blocks, indicating 
that X33 is not a diffuse X-ray emission feature. 
In the case of X10 and X24, time variability is only suggested via a
comparison of the PSPC measured fluxes with the HRI upper limits. While
a PSPC true color picture (calculated from the images in the soft, hard1 and
hard2 band, Vogler 1997) indicates harder spectral behavior
at the position of X10 than for the surrounding diffuse emission, such
a signature is not visible for X24. This might suggest that the X-ray emission
of X10 is due to a background object, whereas that of X24 might
reflect structure in the diffuse halo emission.
No time variability could be found for X45, and therefore this source is most 
likely of a similar origin to X24.

\section{Summary}
The properties of
point-like X-ray sources, detected in deep ROSAT HRI and PSPC observations
of the NGC~253 field, are presented and discussed. Long term
time-variability is established within the ROSAT data and with the help of 
the {\it Einstein} HRI data. Optical follow-up observations of some of the
X-ray sources located (projected) within the halo of NGC~253 
have been carried out at the European Southern Observatory, La Silla, Chile.

\vskip .1cm \noindent $\diamond$ 
The disk of NGC~253 is visible with a count rate of
0.15~cts~s$^{-1}$ in the ROSAT 0.1--2.4~keV HRI band. 
The X-ray emission is a superposition of (1)
individual point sources with luminosities ranging from 
$7 \times 10^{36}$~erg~s$^{-1}$ to $3 \times 10^{38}$~erg~s$^{-1}$ 
(integral count rate of 0.03~HRI~cts~s$^{-1}$ 
corresponding to $L_{\rm x} = 1\times 10^{39}$~erg~s$^{-1}$), (2) an extended
source covering the bulge and connected to the nuclear star forming activity
(0.04~HRI~cts~s$^{-1}$), (3) patchy X-ray emission covering the inner spiral 
arms of NGC~253, (4) diffuse X-ray emission from the disk of NGC~253 and
(5) contributions from the X-ray halo around NGC~253 projected into the
$D_{25}$ ellipse of NGC~253.
These results are in good agreement with the {\it Einstein} and ASCA data. 

\vskip .1cm \noindent $\diamond$ 
Time variability investigations of the ROSAT data suggest that 9 of the
27 NGC~253 disk sources are variable, 2 showing transient behavior.
The {\it Einstein} data establish four additional time-variable sources, 
not seen with ROSAT. Excluding the bright point-like source X33,
located
$\sim 20''$ south of the nucleus, the time-variable sources have maximum
luminosities ranging from $3\times 10^{37}$~erg~s$^{-1}$ to 
$1.5\times 10^{38}$~erg~s$^{-1}$. We suggest that these sources 
are X-ray binaries radiating close to, or at, the Eddington limit of 
an accreting neutron star X-ray binary. If X33 were a binary, 
this source (intrinsic $L_{\rm x} = 4\times 10^{38}$~erg~s$^{-1}$) 
would be a good candidate for a black hole binary.

\vskip .1cm \noindent $\diamond$
A slight X-ray excess 
($L_{\rm x}= (7.4$$\pm$$4.2)\times 10^{36}$~erg~s$^{-1}$) 
at the position of the type I supernova SN 1940 E is detected. However, 
this X-ray emission might be caused by the patchiness of the diffuse
X-ray emission covering the inner spiral arms of NGC~253. 

\vskip .1cm \noindent $\diamond$
The luminosity distribution of the NGC~253 point sources matches those of 
M~31 and M~33. The ratio between the integrated X-ray point-source luminosity
and the optical luminosity ($L_{\rm B}$) is of the order
$4\times 10^{-5}$ after correcting for the different viewing geometries.
In contrast to some more distant spiral galaxies, no very bright 
($L_{\rm x} > 10^{39}$~erg~s$^{-1}$) point-like sources are
detected in NGC~253.

\vskip .1cm \noindent $\diamond$
None of the globular cluster candidates reported from optical observations of
the outer disk and halo of NGC~253 are visible in our X-ray observations, 
though X-ray observations of M~31 suggest
that one half of the NGC~253 sources should be due to globular
cluster X-ray binaries. This non-detection might be a problem caused by 
the very small number of optical candidates located close ($d<5$~kpc) 
to the plane of the galaxy. 

\vskip .1cm \noindent $\diamond$ The NGC~253 halo shows filamentary diffuse
X-ray emission. 7 point like sources are detected in the region of
diffuse X-ray emission. For 4 point sources time variability or proposed
optical identifications rule out the idea 
that these sources are connected to the 
diffuse halo emission.

\begin{acknowledgements}
We thank our colleagues from the MPE ROSAT group for their support.
The ROSAT project is supported by the German Bundesministerium
f\"ur Bildung, Wissenschaft, Forschung und Technologie (BMBF/ DLR) 
and by the Max-Planck-Gesellschaft. We thank Andrew Read and
Stefan Immler for carefully reading and discussing the manuscript.
To look for possible counterparts of our point source list we made use of
the NASA/IPAC extragalactic database (NED). 
The usage of the likelihood ratio 
test for time variability studies of sources with low photon statistics
(cf. Sect.~\ref{timevardet}) was proposed by M.G. Akritas on the
WWW homepage of the Statistical Consulting Center for Astronomy
(http://www.stat.psu.edu/scca/homepage.html). 
\end{acknowledgements}

\begin{appendix}

\section{Sources located outside the area covered by the $D_{25}$ ellipse of
         NGC~253}

\begin{figure}
\label{wnpspectra}
\unitlength=1cm
\begin{picture}(10,12)
\put(0,8){\includegraphics[bb=100 410 660 626,width=8.8cm,height=4cm,clip=true]{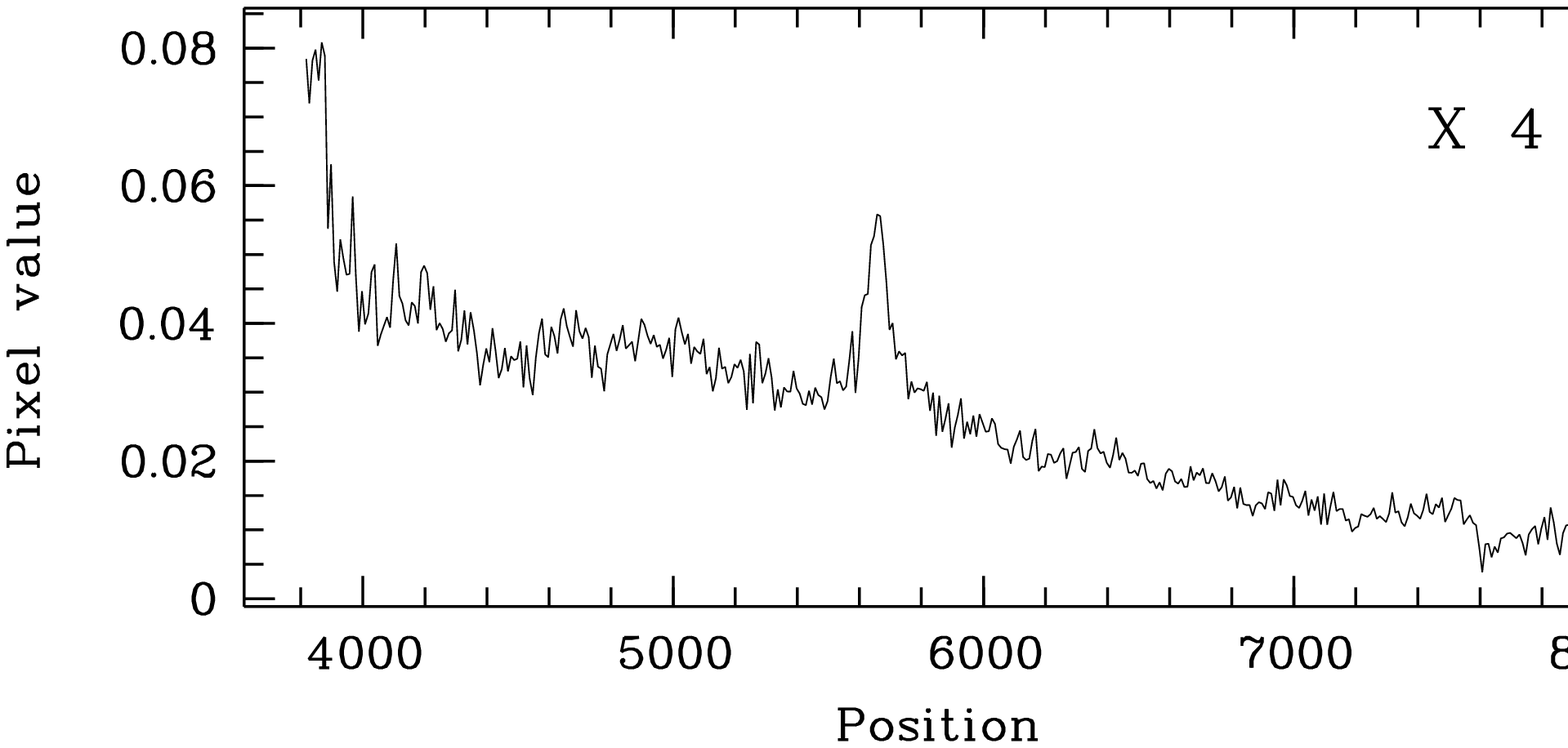}}
\put(0,4){\includegraphics[bb=100 410 660 626,width=8.8cm,height=4cm,clip=true]{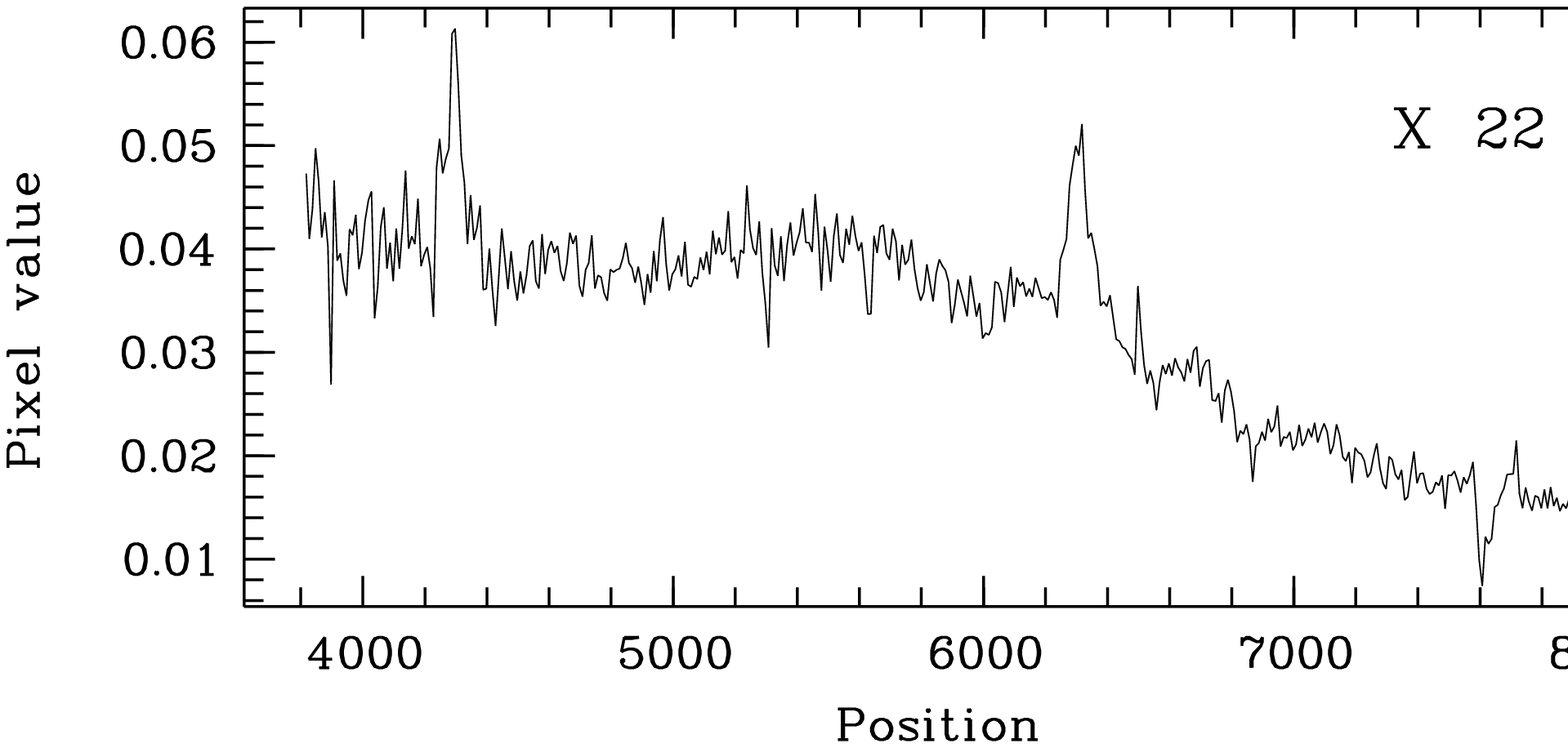}}
\put(0,0){\includegraphics[bb=100 387 660 626,width=8.8cm,height=4cm,clip=true]{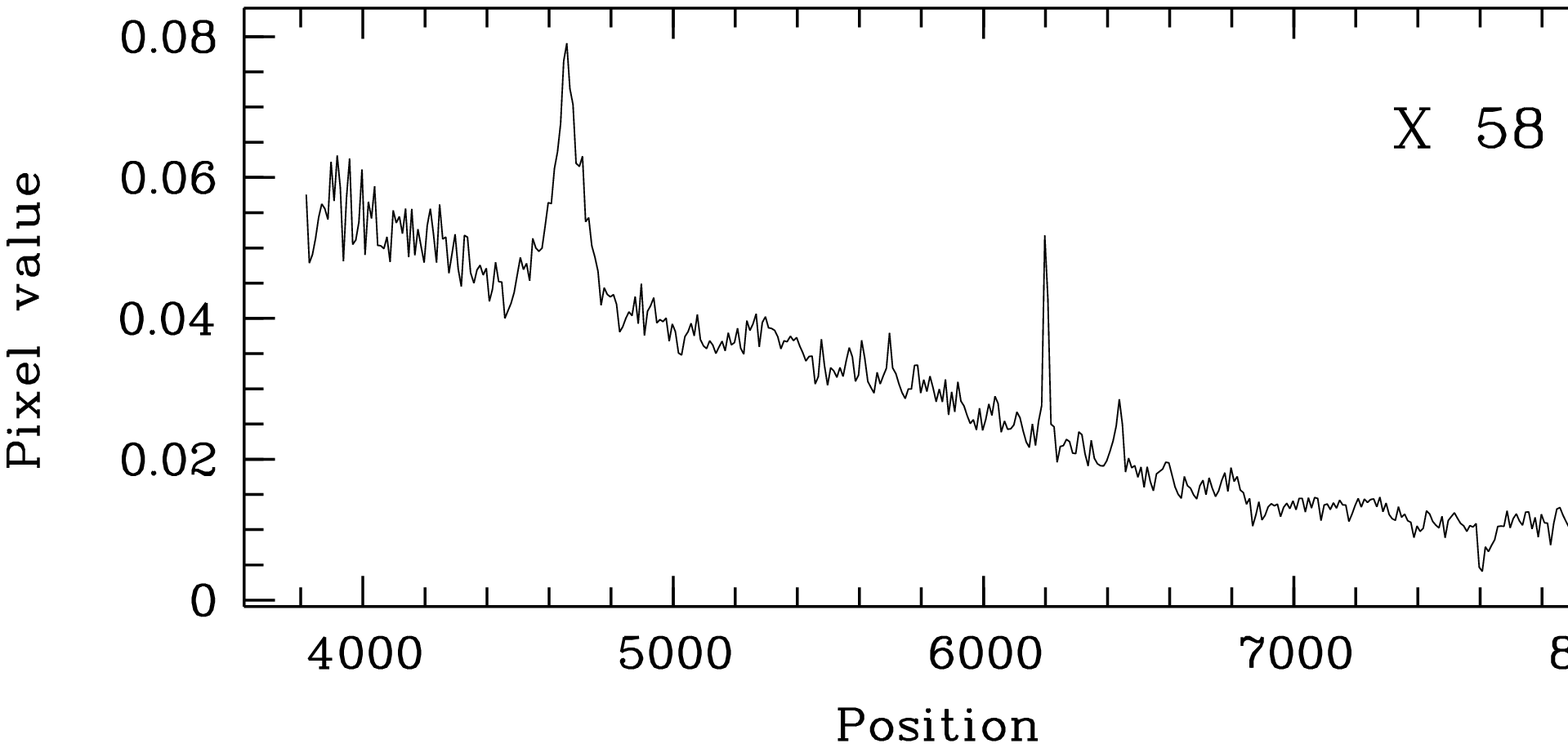}}
\end{picture}
\caption{
          Optical spectra of QSO identification candidates for X4, X22, X58.
          $f_\lambda$ in units of $10^{-15}$~erg~cm$^{-2}$~s$^{-1}$~\AA$^{-1}$
          is plotted against wavelength in \AA
        }
\end{figure}

%
%====================================================
% Spalten 1,2,3,4,5,7,10 aus der Miadstabelle finding_98.tbl
% We, 21 Jan 1998   16:49:23
%====================================================
%

\begin{table*}
\caption{Sources located outside the area covered by the $D_{25}$ ellipse of
         NGC~253}
\label{nolabel}
\begin{flushleft}
\begin{tabular}{rrrrrcccl}
\hline
\noalign{\smallskip}
 &
$\alpha_{2000}^{\rm ROE~ object}$ &
$\delta_{2000}^{\rm ROE~ object}$ &
dist. &  $^\ddagger$  
& blue &
$L_{\rm x}^{\rm HRI}/L_{\rm B}$ & $L_{\rm x}^{\rm PSPC}/L_{\rm B}$ 
& comments, results of ESO\\
& & & $''$~~ & & mag&$^\star$ &$^\star$ & follow-up observations\\
\noalign{\smallskip}
\hline
\noalign{\smallskip}
%
% < ******* now row 01 is following ******* >
%
X1 &
00 46 12.41 &
$-$25 16 59.5 &
$1.7$ &
 s &
$19.96$ & & 3.8e$-$1
                                                   \\
& 00 46 12.29 & $-$25 16 49.1 & 9.6 & f & 22.67 & & 4.5e$+$0\\
%
% < ******* now row 03 is following ******* >
%
X3 &
00 46 42.94 &
$-$25 38 17.9 &
$6.4$ &
 s &
$15.39$ & & 1.3e$-$2
                                                                                 \\
%
% < ******* now row 04 is following ******* >
%
X4 &
00 46 47.28 &
$-$25 21 50.8 &
$2.0$ &
 s &
$20.26$ & 3.4e$-$1 & 3.1e$-$1 &
QSO, $z=1.022\pm 0.001$                                              \\
%
% < ******* now row 13 is following ******* >
%
X13 &
00 47 09.47 &
$-$25 14 04.4 &
$0.6$ &
 f &
$23.72$& 3.9e$+0$ &  4.7e$+0$ & \\%{\bf star picked up by ESO}\\
%
% < ******* now row 22 is following ******* >
%
X22 &
00 47 22.98 &
$-$25 10 54.9 &
$1.9$ &
 s &
$19.94$ & 1.8e$-$1 &  1.7e$-$1&
QSO, $z=1.250\pm 0.003$ \\
%
% < ******* now row 27 is following ******* >
%
X27 &
00 47 30.12 &
$-$25 08 59.5 &
$11.6$ &
 f &
$22.20$ & & 1.1e$+0$
                                                                                 \\
%
% < ******* now row 31 is following ******* >
%
X31 &
00 47 32.18 &
$-$25 28 12.0 &
$2.9$ &
 s &
$11.38$  &5.3e$-$5 &  5.4e$-$5 & %{\bf SIMBAD, Hipparcos?}
        \\
%
% < ******* now row 37 is following ******* >
%
X37 &
00 47 38.29 &
$-$25 38 34.9 &
$11.6$ &
 s &
$21.19$ & & 2.5e$-$1            \\
%
% < ******* now row 39 is following ******* >
%
X39 &00 47 39.37 & $-$25 25 33.9 & 14.3 & f & 23.01& & 7.3e$-$1\\
%
% < ******* now row 43 is following ******* >
%
X43 &
00 47 44.23 &
$-$25 26 51.5 &
$4.0$ &
 s &
$21.11$ & 1.7e$-$1 &  1.1e$-$1
                                                                                 \\
%
% < ******* now row 46 is following ******* >
%
X46 &
00 47 46.55 &
$-$25 27 36.0 &
$3.5$ &
 f &
$21.74$ & 4.1e$-$1 &  2.1e$-$1
                                                                                 \\
%
% < ******* now row 47 is following ******* >
% 
X47 &
00 47 46.55 &
$-$25 29 55.7 &
$3.3$ &
 s &
$19.68$ & 1.1e$-$1 &  1.5e$-$1
                                                                                 \\
%
% < ******* now row 49 is following ******* >
%
X49 & 00 47 50.04 & $-$25 00 34.6 & $^\dagger 17.3$ & f & 22.84 & 4.7e$+$0
& 2.3e$+$0\\  
%
% < ******* now row 53 is following ******* >
%
X53 &
00 47 50.33 &
$-$25 08 42.1 &
$2.3$ &
 f &
$22.06$ & 3.8e$-$1  
                                                                                 \\
%
% < ******* now row 54 is following ******* >
%
X54 &
00 47 51.27 &
$-$25 03 24.5 &
$4.6$ &
 f &
$21.38$ &4.8e$-$1 &  5.6e$-$1

                                                                                 \\
%
% < ******* now row 58 is following ******* >
%
X58 &
00 48 00.04 &
$-$25 09 53.8 &
$3.3$ &
 s &
$18.53$ & 5.8e$-$2  & 6.7e$-$2 &
QSO, $z=0.664\pm0.001$
                                                     \\
%
% < ******* now row 59 is following ******* >
%
X59 &
00 48 01.18 &
$-$25 27 38.4 &
$8.3$ &
 g &
$18.14$ & & 9.1e$-$3  \\
& 00 48 01.26 & $-$25 27 31.4 & 10.0 & s & 20.11 & &5.8e$-$1 \\
%
% < ******* now row 60 is following ******* >
%
X60 &
00 48 00.93 &
$-$25 23 51.9 &
$2.4$ &
 f &
$22.69$ & 2.4e$+0$  & 2.3e$+0$  \\
%
% < ******* now row 61 is following ******* >
%
X61 &
00 48 02.83 &
$-$25 04 44.1 &
$^\dagger 12.6$ &
 s &
$\phantom{0}8.90$ &4.0e$-$6 &  3.9e$-$6 & %{\bf SIMBAD, Hipparcos?}
\\                        
%
% < ******* now row 63 is following ******* >
%
X63 &
00 48 04.40 &
$-$25 06 04.7 &
$3.1$ &
 s &
$19.45$ & & 1.8e$-$2
                                                                                 \\
%
% < ******* now row 64 is following ******* >
%
X64 &
00 48 08.46 &
$-$25 25 07.2 &
$15.8$ &
 f &
$23.78$ & & 1.6e$+0$
                                                                                 \\
%
% < ******* now row 66 is following ******* >
%
X66 &
00 48 09.05 &
$-$25 04 53.6 &
$3.5$ &
 f &
$21.74$ &  1.0e$+0$ &  1.4e$+0$
                                                                                 \\
%
% < ******* now row 68 is following ******* >
%
X68 &
00 48 29.87 &
$-$25 08 25.9 &
$^\dagger 14.9$&
 f &
$22.15$             &        1.5e$+0$  & 2.3e$+0$                                         \\
%
% < ******* now row 69 is following ******* >
%
X69 &
00 48 31.90 &
$-$25 15 19.9 &
$5.9$ &
 f &
$23.01$ & &5.6e$-$1
                                                                                 \\
%
% < ******* now row 70 is following ******* >
%
X70 &
00 48 43.91 &
$-$25 29 40.5 &
$9.6$ &
 s &
$19.41$ & & 7.7e$-$2
                                                                                 \\
%
% < ******* now row 71 is following ******* >
%
X71 &
00 48 47.47 &
$-$25 21 58.3 &
$^\dagger 13.5$ &
 s &
$20.77$ & & 3.9e$-$1 & X71 and X72 are located at  the\\
%
% < ******* now row 72 is following ******* >
%
X72 &
00 48 47.92 &
$-$25 07 53.2 &
$12.0$ &
 s &
$20.13$ & & 1.8e$-$1 &
position of the PSPC support ring\\
%
% < ******* now row 73 is following ******* >
%
X73 &
00 48 58.78 &
$-$25 00 17.4 &
$8.2$ &
 s &
$13.25$ & & 1.1e$-$3
                                                                                 \\
\noalign{\smallskip}
\hline
\end{tabular}
\end{flushleft}
\vskip-.5cm
\[
\begin{array}{lp{0.95\linewidth}}
^\ddagger & Object classification according to the ROE charts: f = faint,
s = stellar, g = galaxy\\
^\star & e$-1$ means $\times 10^{-1}$ etc.
\\
^{\dagger} & Separation of ROE source and X-ray source exceeds the 
             error radius of the X-ray source by a few arcsec \\
\end{array}
\]
\end{table*}

The X-ray properties and light curves of sources located outside the
$D_{25}$ ellipse of NGC~253 have already been presented in Table~3 and Fig.~5.
In this appendix proposed optical identifications of the sources
obtained from the ROE finding charts (Irwin et al. 1994)
are listed (Table~\ref{nolabel}). For all sources, the
ratio of the X-ray luminosity to the optical luminosity ($L_{\rm B}$) 
is given.
From observations of stars in our Galaxy (Neuh\"auser 1998, priv. comm.) one
expects ratios of $L_{\rm x}/L_{\rm B}
\la 3\times 10^{-3}$. For that reason, only the
X-ray emission of the sources X31, X61 and possibly X73 can be explained
by the detection of stars, while the other sources are most likely 
extragalactic objects.

None of the X-ray field sources correlated with entries in the 
NASA IPAC extragalactic database (NED). However,
X4 and X70 coincide with members of a list of quasar 
candidates (Crampton et al. 1997) and X59 with a radio point source 
(see radio images of Hummel et al. (1984) and NVSS catalog by Condon et al. 
(1998)), giving further support to 
their identification as background objects.    

From November 5th to 7th, 1996, during an observation campaign at the 2.2~m
ESO/MPG telescope at La Silla observatory, spectra of some of the
optical candidates were obtained using the EFOSC2 spectrograph
with grism \#6 and a 1\farcs5 wide long slit and the 2048$\times$2048
15~$\mu$m LORAL CCD, which gives a dispersion of 2\AA\ per pixel,
a spectral coverage of 3800~\AA\ -- 8000~\AA, and a spectral resolution
of 10~\AA. The seeing typically was between 0\farcs8 and 1\farcs5.
The data were reduced according to Pietsch et al. 1998c.
Spectra of the optical candidates for X4, X22 and X58 (Fig.~A1) 
clearly identified them as Quasars with redshifts of 0.6 to 1.3 (see 
Table~11) using C{\sc iii} $\lambda\,1909$, Mg\,{\sc ii} $\lambda\,2798$,
[O {\sc ii}] $\lambda\,3727$ [Ne {\sc iii}] $\lambda\,3869,~3968$ lines.
As regards X13, we only took a spectrum of a
stellar ROE object with B mag 18.5 close (7$''$) to X13, though clearly outside
the error circle, the stellar object close by being to faint for spectroscopic
follow-up with this instrumentation. By comparing with the library of stellar spectra
of Jacoby et al. (1984), the object can be classified as a M0\,V star. The
large offset and the $L_{\rm x}/L_{\rm B}$ ratio of 0.04 however 
-- extremely high for a star of this type -- strongly argue against this
identification with X13.

\end{appendix}

%                         pec: A=absence of bar B=Bar X=intermediate
% galaxy          TType   pec  Seyf  incl  Vtype  d     N_H  ref
% 
% 253             5 Sc               86           2.58  1.3  this work  
% 4258            4 Sbc   X    L/S   71           6.4   1.2  Vogler 1997
% M31 = 224       3 Sb    A          78           0.69  3.0  Supper et al. 1997
% M51 = 5194      4 Sbc   AP   L/S   64     7.7   7.7   1.3  Immler 1996
% M81 = 3031      2 Sab   A    L1.9  60           3.5   3.8  Fabbiano 1988
% M100 = 4321     4 Sbc   X          37           17.1  2.3  Immler et al. 1998
% M101 = 5457     6 Scd   X          0      7.5   7.5   1.1  Immler 1996

%                 number of ps  integral lum of ps
%                 vis with det. without nucl/bulge  L_x/L_B 
% 253             25 + nucl     1.0e39              0.0000223
% 4258            13 + nucl     2.8e39              0.0000307
% M31 = 224       340+ bulge    1.8e39              0.0000168
% M51 = 5194      17 + nucl     7.9e39              0.0000431
% M81 = 3031       8 + nucl     2.5e39              0.0000323
% M100 = 4321      9 + nucl     7.3e39              0.0000421
% M101 = 5457     21 + nucl     6.0e39              0.0000346


\begin{thebibliography}{}

\bibitem[ ] { } Barbon R., Capellaro E., Turatto M., 1989, A\&AS 81, 421
\bibitem[ ] { } Beck R., Hutschenreiter G., Wielebinski R., 1982, A\&A 106, 
                112
\bibitem[ ] { } Beck R., Carilli C.L., Holdaway M.A., et al., 1994, 
                A\&A 292, 409
\bibitem[ ] { } Blair W.P., Kirshner R.P., Winkler P.F., 1983, ApJ 272, 84
\bibitem[ ] { } Blecha A., 1986, A\&A 154, 321
\bibitem[ ] { } Bregman J.N., Pildis R.A., 1994, ApJ 420, 570
\bibitem[ ] { } Cash W., 1979, ApJ 228, 939
\bibitem[ ] { } Carilli C.L., Holdaway M.A., Ho P.T.P., et al.,
                1992, ApJ 399, L59
\bibitem[ ] { } Carral P., Hollenbach D.J., Lord S.D., et al.,
                1994, ApJ 423, 223 
\bibitem[ ] { } Chu Y.H., Kennicutt R.C., 1994, ApJ 425, 720
\bibitem[ ] { } Condon J.J., Cotton W.D., Greisen E.W., et al., 1998, 
                AJ 115,1693
\bibitem[ ] { } Crampton D., Gussie G., Cowley A.P., Schmidtke P.C., 1997, 
                AJ 114, 2353
\bibitem[ ] { } Eddington A.S., 1928, ``The Internal Constitution of Stars'' 
                (1928), Dover (New York 1959)
%\bibitem[ ] { } Ehle M., Pietsch W., Beck R., Klein U., 1998, A\&A 329, 39
\bibitem[ ] { } ESO MIDAS 1997,
       http://www.eso.org/research/data-man/data-proc/systems/esomidas/
\bibitem[ ] { } Gehrels N., 1986, ApJ 303, 336
\bibitem[ ] { } Fabbiano G., 1988, ApJ 330, 672
%\bibitem[ ] { } Fabbiano G., 1988b, ApJ 325, 544
\bibitem[ ] { } Fabbiano G., Trinchieri G., 1984, ApJ 286, 491
\bibitem[ ] { } Fabbiano G., Kim D.-W., Trinchieri G., 1992, ApJS 80, 531
\bibitem[ ] { } Fabian A.C., Terlevich R., 1996, MNRAS 280, L5
\bibitem[ ] { } Forbes D.A., Ward M.J., Depoy D. L., 1991, ApJ 380, 63
%\bibitem[ ] { } Immler S., diploma thesis, Ludwig Maximiliam Universit\"at,
%                Munich
\bibitem[ ] { } Hasinger G., Schmidt M., Tr\"umper J., 1991, A\&A 246, L2
\bibitem[ ] { } Hasinger G., Burg R., Giacconi R., et al., 1993, A\&A 275, 1
\bibitem[ ] { } Hogg R.V., Tanis E.A., 1983, Probability and statistical 
                inference, Macmillan
\bibitem[ ] { } Hummel E., Smith P., van der Hulst J.M., 1984, A\&A 137, 138
\bibitem[ ] { } Immler S., Pietsch W., Aschenbach B., 1998a, A\&A 331, 601 
\bibitem[ ] { } Immler S., Pietsch W., Aschenbach B., 1998b, A\&A 336, L1
\bibitem[ ] { } Irwin M., Maddox S., McMahon R., 1994, Spectrum 2, 14
\bibitem[ ] { } Jacoby G.H., Hunter D.E., Christian C.A., 1984, ApJS 56, 257
%\bibitem[ ] { } Kowal C.T., Sargent W.L.W., 1971, ApJ 76, 756
\bibitem[ ] { } Liller W., Alcaino G., 1983, ApJ 265, 166
\bibitem[ ] { } Pfeffermann E., Briel U.G., Hippmann H., et al., 1987, 
                presented at SPIE, Berlin, West Germany 1986, cf.
                Proc. SPIE 733, 519
\bibitem[ ] { } Pietsch W., et al., 1998a, in prep. 
\bibitem[ ] { } Pietsch W., Trinchieri, G., Vogler, A., 1998b, A\&A, submitted
\bibitem[ ] { } Pietsch W., Bischoff, K., Boller Th., et al., 
                1998c, A\&A 333, 48
\bibitem[ ] { } Primini F.A., Forman W., Jones C., 1993, ApJ 410, 615
\bibitem[ ] { } Ptak A., Serlemitsos P., Yaquoob T., 
                Mushotzky R., Tsuru T., 1997, AJ 113, 1286
\bibitem[ ] { } Puche D., Carignan C., van Gorkom J.H., 1991, AJ 101, 2
\bibitem[ ] { } Raymond J.C., Cox D.P., Smith B.W., 1976, ApJ 204, 290 
\bibitem[ ] { } Read A.M., Ponman T.J., Strickland D.K., 
                1997, MNRAS 286, 626
\bibitem[ ] { } Sams III B.J., Genzel R., Eckhart A., et al., Tacconi-Garman 
                L., Hofmann R., 1994, ApJ 430, L33
\bibitem[ ] { } Schlegel E.M., 1994, ApJ 434, 523
\bibitem[ ] { } Schlegel E.M., 1995, Rep. Prog. Phys. 58, 1375
\bibitem[ ] { } Schulman E., Bregman J.N., 1995, ApJ 441, 568
\bibitem[ ] { } Snowden S., McCammon D., Burrows D., et al., 
                1994, ApJ 424, 714
\bibitem[ ] { } Sofue Y., Wakamatsu K.-I., Malin D.F., 1994, AJ 108, 2102
\bibitem[ ] { } Suchkov A.A., Dinshaw S.B., Heckman T.M., Leitherer C.,
                1994, ApJ 430, 511
\bibitem[ ] { } Supper, R., Hasinger, G., Pietsch, W., et al.,
                1997, A\&A 317, 328
\bibitem[ ] { } Trinchieri G., Fabbiano G., Peres G., 1988, ApJ 325, 531
\bibitem[ ] { } Tr\"umper J., 1983, Adv. Space Res. 2, 241
\bibitem[ ] { } Ulvestad J.S., Antonucci R.R.J., 1997, ApJ 488, 621
\bibitem[ ] { } van den Bergh S., 1993, Comments on Astrophys. 17, 125
\bibitem[ ] { } Voges W., Gruber R., Paul J., et al., 1992,
                ``The ROSAT Standard Analysis Software System''. 
                In: Proc. European ISY meeting, Symposium ``Space Sciences 
                with Particular Emphasis on High Energy Astrophysics'', p. 223
\bibitem[ ] { } Vogler A., 1997, PhD thesis
\bibitem[ ] { } Vogler A., Pietsch W., 1997, A\&A 319, 459 
\bibitem[ ] { } Vogler A., Pietsch W., Kahabka P., 1996, A\&A  305, 74
\bibitem[ ] { } Vogler A., Pietsch W., Bertoldi F., 1997, A\&A 318, 768
\bibitem[ ] { } Waller W.H., Kleinmann S.G., Ricker G.R., 1988, AJ 95, 1057
\bibitem[ ] { } Watson A.M., Gallagher III J.S., Holtzman J.A., et al., 
                1996, AJ 112, 534
\bibitem[ ] { } Williams R.M., Chu Y.H., 1995, ApJ 439, 132
\bibitem[ ] { } Zimmermann H.U., Lewin W., Predehl P., et al., 1994,
                Nat 367, 621
\bibitem[ ] { } Zimmermann H.U., Becker W., Belloni T., et al., 1997, 
                ``EXSAS users guide'', July 1997 edition 

\end{thebibliography}
\end{document}